\documentclass[graybox, envcountchap]{svmult}

\usepackage{mathptmx}        
\usepackage{helvet}          
\usepackage{courier}         

\usepackage{makeidx}         
\usepackage{graphicx}        
\usepackage{multicol}        
\usepackage[bottom]{footmisc}

\makeindex             

\usepackage{amsmath}
\usepackage{amssymb}
\usepackage{lscape}
\usepackage{multirow}

\def\gapp{\ifmmode\stackrel{>}{_{\sim}}\else$\stackrel{<}{_{\sim}}$\fi}
\def\gsim{\lower.5ex\hbox{\gtsima}}
\def\gtsima{$\; \buildrel > \over \sim \;$}
\def\lapp{\ifmmode\stackrel{<}{_{\sim}}\else$\stackrel{<}{_{\sim}}$\fi}
\def\lsim{\lower.5ex\hbox{\ltsima}}
\def\ltsima{$\; \buildrel < \over \sim \;$}

\newcommand\apgt{\ {\raise-.5ex\hbox{$\buildrel>\over\sim$}}\ }
\newcommand\aplt{\ {\raise-.5ex\hbox{$\buildrel<\over\sim$}}\ }
\begin{document}
\pagestyle{empty}
\frontmatter

\include{dedic}
\include{foreword}
\include{preface}

\mainmatter

\setcounter{chapter}{5}

\title{The Blue Straggler Population in Dwarf Galaxies}
\author{Yazan Momany}
\institute{Yazan Momany \at European Southern Observatory, Alonso de Cordova 3107, Santiago,
Chile, and\\ 
INAF, Oss. Astronomico di Padova, Vicolo dell'Osservatorio 5, I-35122 Padova, Italy,\\
 \email{ymomany@eso.org}}

%
%
\maketitle
\label{Chapter:Momany}

\abstract*{In this chapter I review the recent developments regarding the
study of Blue Stragglers (BSS) in dwarf galaxies. The loose density
environment of dwarf galaxies resembles that of the Galactic Halo,
hence it is natural to compare their common BSS properties. At the
same time, it is unescapable to compare with the BSS properties in
Galactic globular clusters, which constitute \emph{the} \emph{reference
point} for BSS studies. Admittedly, the literature on BSS in dwarf
galaxies is not plentiful. The limitation is mostly due to the large
distance to even the closest dwarf galaxies. Nevertheless, recent
studies have allowed a deeper insight on the BSS photometric properties
that are worth examining.}

\section{Introduction}
\label{mom:1}
\subsection{Blue Stragglers in Globular Clusters}

Ever since their first identification in M3\index{M3} by Sandage \cite{YoSandage1953},
the term \emph{Blue Stragglers} (BSS) refers to the population of stars
that are relatively brighter and bluer than the cluster's main sequence\index{main sequence}
(MS) turn-off\index{turn-off} point. Several decades after, globular clusters\index{globular cluster} remain
the ideal environment where BSS can be identified with certain \emph{ease}.
Indeed, the family of Galactic globular clusters typically shows a
turn-off mass of $\sim0.8$ M$_{\odot}$, being mostly coeval
and $\gtrsim10$ Gyr old \cite{Marin-Franch2009}. Thus, the
identification of a $\sim1.2\div1.5$ M$_{\odot}$ hot stellar population,
\textit{nowadays}, is unexpected since such high mass stars would
have evolved already in a $\gtrsim10$ Gyr system; thereafter the
\textit{``straggler''} term. 

In the context of globular clusters studies the last decade has brought
a new wave of interest in these presumably \emph{dead systems} (see Chap. 5). In
particular, thanks to the accumulating evidence of: (i) systematic
chemical abundance anomalies\index{chemical anomaly} \cite{Carretta2009f}; coupled with
(ii) detection of multiple and discrete red giant\index{red giant} branches and/or
main sequences \cite{Milone2013d}, the idea that multiple stellar
populations\index{stellar
population} coexist within \emph{a} single cluster is taking the upper
hand \cite{Renzini2013}. This clearly defies the old text-book definition
of globular clusters (i.e. \emph{ancient, coeval and chemically homogeneous
systems}).

For what concerns the BSS identification in globular clusters one
may wonder if admitting the multiple populations scenario would leave
some space for the presence of a ``young'' population (BSS), the
answer is \emph{no}. Indeed, and based on our current understanding,
the complexity of multiple epochs of star formation within globular
clusters is always limited to the first $\lesssim1$ Gyr since the
formation epoch of the cluster (some $\sim13$ Gyr time ago). Thus,
and despite these chemical anomalies and multiple generations of stars,
there is still no room to accommodate for the presence of a $1.2\div1.5$
M$_{\odot}$ hot/blue stellar population (i.e BSS) within the \emph{standard
single-star} evolution in globular clusters.

The origin of the BSS is usually sought as either the products of
(i) dynamical interactions\index{dynamical interaction}; or (ii) binary evolution\index{mass transfer} (see Chap. 6). The dynamical
origin for BSS foresees a \emph{continuous} production of collisional\index{collision}
binaries \cite{Hills1976d} between single and/or binary MS stars\index{binary star}
throughout the life of the stellar system. On the other hand, the
most likely BSS formation scenario is that involving binary evolution.
In this case the origin of the fresh hydrogen that ``rejuvenates''
the BSS is mass transfer\index{mass transfer} \cite{McCrea1964a}. In particular,
mass transfer occurs when the evolved primary (now invisible) fills
its Roche lobe\index{Roche lobe} and processed material overflows to the secondary MS
(which \emph{now} constitutes the BSS visible component). Within the
binary evolution scenario, BSS can also be formed via the coalescence
of a binary system made by two ``normal'' MS stars at the TO level.

\subsection{The Importance of Dwarf Galaxies}

Dwarf galaxies\index{dwarf galaxy} represent the dominant population, by number, of the
present-day universe and galaxy clusters. These low-mass galaxies
are \emph{not} simply scaled-down versions of giant systems. Indeed,
they hold the keys for a deeper understanding of galaxy formation,
chemical evolution, star formation processes and dark matter content.
In the framework of hierarchical clustering scenarios such as in cold
dark matter models \cite{Blumenthal1984d} dwarf galaxies would
have been the first objects to be formed, that would later contribute
to the assembling of larger systems \cite{White1978e}. The resultant
picture is one in which the dwarfs observed nowadays are those which
survived merging events. There are a number of factors that contribute
to making the Local Group\index{Local Group galaxy} ($d\lesssim1.1$ Mpc) a unique laboratory
for astronomers; first its dwarf members are close enough to enable
determining age, metallicity\index{metallicity} and their star formation histories from
their \textit{resolved stellar populations}. Second, the Local Group
comprises dwarf galaxies of such variety of morphological types, masses,
ages, spatial distributions and metallicities that they are statistically
representative of other dwarf populations present in other environments,
nearby groups or clusters. The two aforementioned properties qualify
dwarf galaxies as optimal cases for ``\textit{near-field cosmology}''.

When studying the Blue Stragglers properties in the resolved stellar
populations of nearby dwarf galaxies one soon realises a basic limitation:
a meaningful analysis is possible only for those galaxies whose imaging
surveys reach at least $\sim$1 magnitude below the old main-sequence
turnoff level. Consequently, the current sample of ``\textit{useful}''
Local Group dwarf galaxies is limited to systems within $D_{\odot}\approx900$
kpc, where the upper limit is set by the Tucana \textit{dwarf spheroidal}
galaxy survey by the \emph{Hubble Space Telescope}\index{Hubble Space Telescope}. Nevertheless, one should
appreciate that addressing the BSS population in such a distant system
implies availing a photometric catalogue with a reasonable photometric\index{photometry}
completeness level at $I\thickapprox29.0$, which is a major challenge
already.

\begin{figure}
\begin{centering}
\includegraphics[width=119mm]{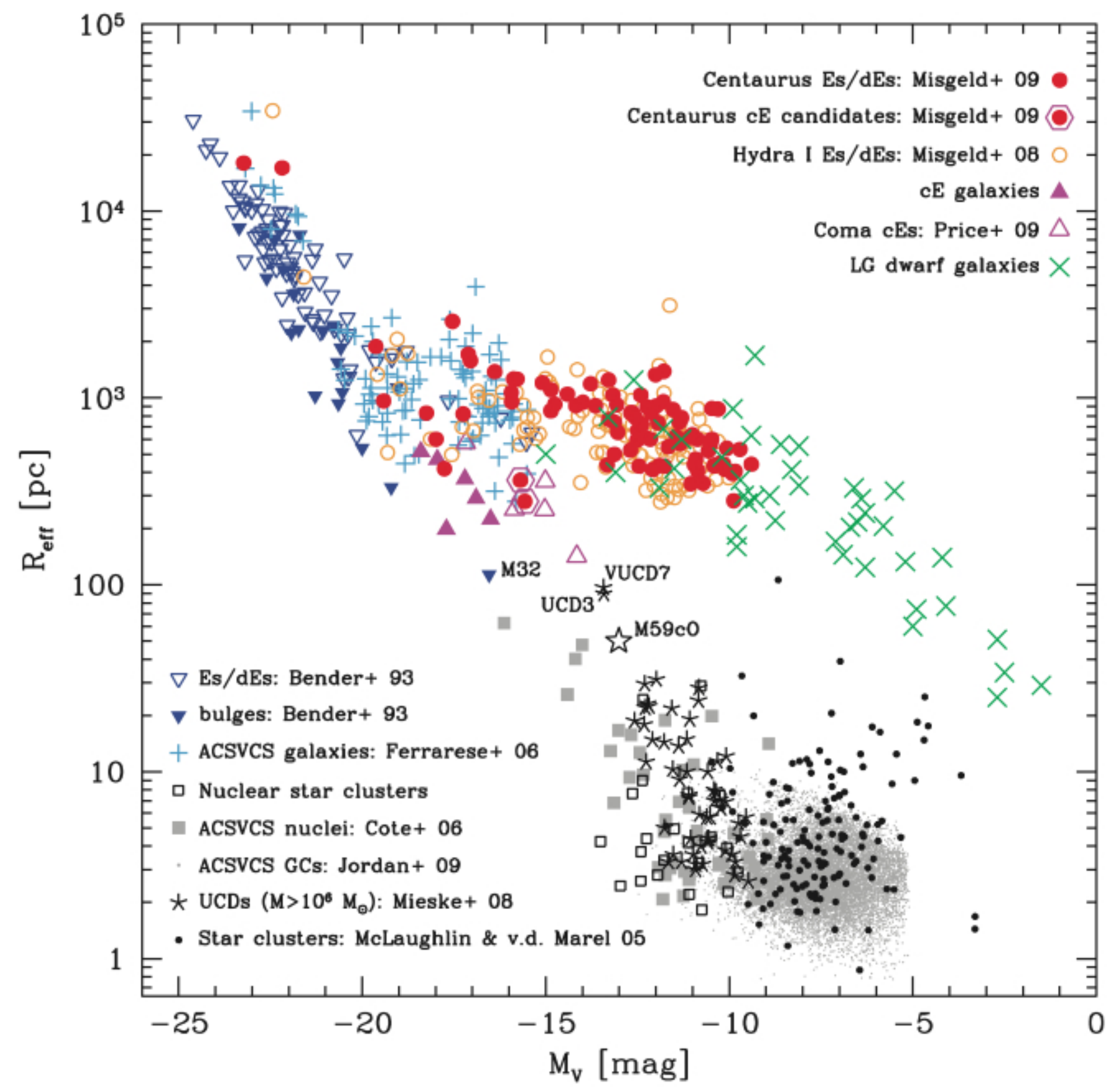}
\par\end{centering}
\caption{The effective radius ($R_{eff}${[}pc{]}) \emph{vs} absolute $V$-band
magnitude ($M_{V}$) for the family of stellar systems covering a
range of 10 magnitudes in mass. This figure is reproduced from \cite{Misgeld2011}
with permission by the RAS.
\label{momfig:Effective-radius-[pc]} }
\end{figure}

\subsection{Dwarf Galaxies vs Globular Clusters}

It is important to recall that the \emph{fuzzy}, almost featureless
dwarf galaxies still represent a different class of objects on their
own, which is fundamentally different from that of globular clusters.
{Misgeld} \& {Hilker} \cite{Misgeld2011} searched for \emph{Fundamental Plane} relations for
systems ranging from faint galaxies and star clusters\index{star cluster} of only a few
hundred solar masses up to giant ellipticals\index{giant elliptical galaxy} $10^{12}$ M$_{\odot}$.
Their analysis (see Fig.\ref{momfig:Effective-radius-[pc]}) shows a
clear dichotomy between the \emph{galaxy} and \emph{star cluster} family. It is straightforward to qualify globular clusters as members
of the \emph{star cluster} family, and that dwarf galaxies occupy
the faint tail of the galaxy group. In particular, and over several
orders of magnitudes ($-10\lesssim M_{V}\lesssim-5$), the typical
effective radius of globular clusters ($\sim3-10$ pc) and dwarf galaxies
(few hundreds to $\sim1000$ pc) does not vary significantly with
mass.

One may argue that the faintest dwarf galaxies reach low effective
radii that are \emph{too close} to typical values of the clusters
group, thereby defying the \emph{galaxy }definition. Clearly, these
are not \emph{ordinary} dwarf galaxies (e.g. Segue$\,$I\index{Segue I} has $M_{V}\sim-1.5$
and a $M/L\sim3000$) rather they probably represent a class of objects
losing dynamical equilibrium and close to disruption \cite{Gilmore2007cm,Niederste-Ostholt2009}. Nevertheless, these ultra-faint
dwarfs (whose total luminosities can be less than that of individual
red giants) show kinematic and metallicity evidence that unambiguously
support the dwarf galaxy classification \cite{Willman2012} and as
such their BSS population (when present) should be included in any
review. This is particularly, important because the sample of ultra-faint
dwarfs is expected to rise thanks to forthcoming Southern hemisphere
surveys (e.g. \emph{Skymapper}\index{Skymapper}%
\footnote{\url{http://rsaa.anu.edu.au/observatories/siding-spring-observatory/telescopes/skymapper}} \cite{Keller2012am}
).

\section{BSS Identification in Dwarf Galaxies}

\begin{figure}
\begin{centering}
\includegraphics[width=119mm]{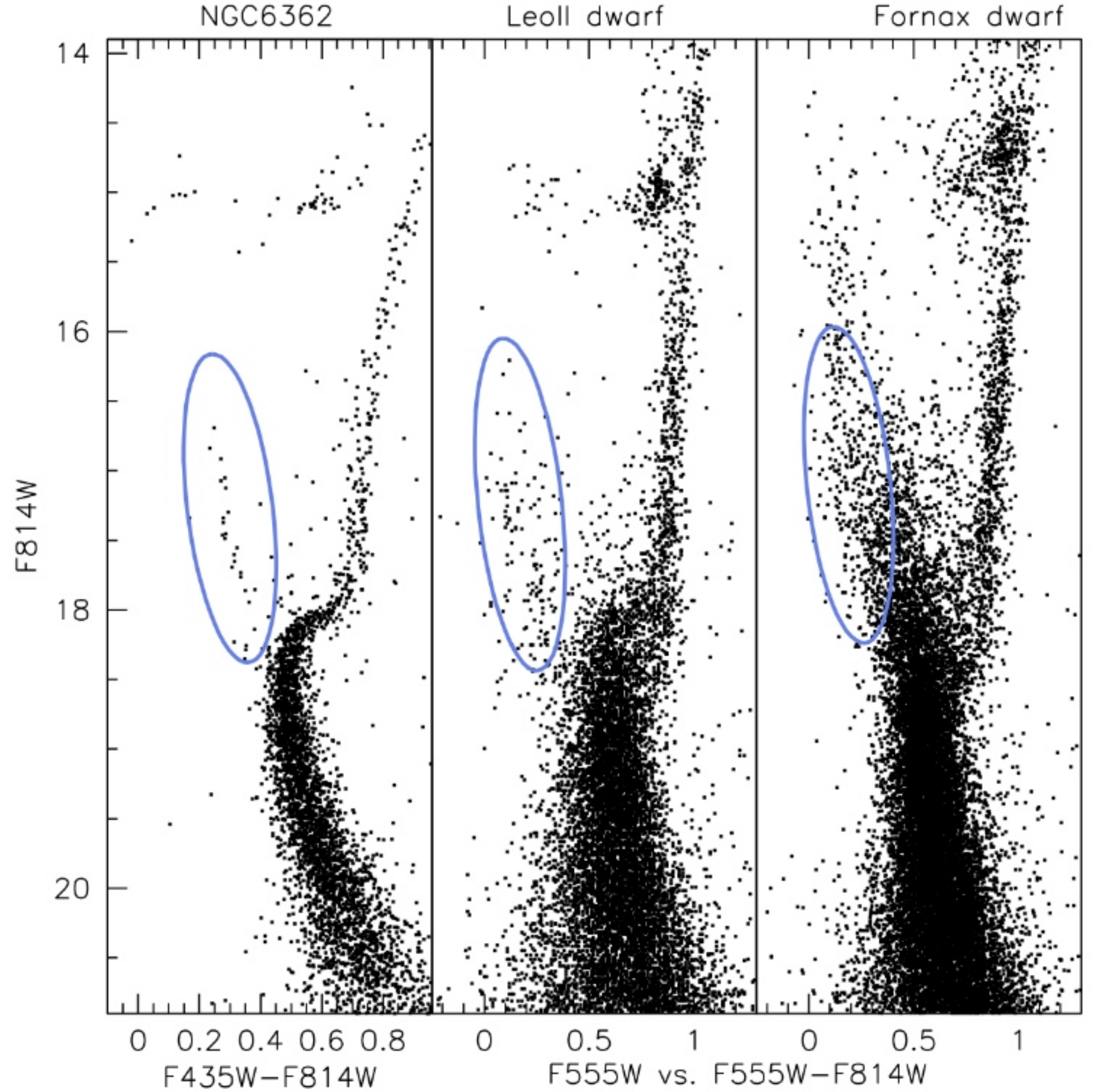}
\par\end{centering}

\caption{From left to right the panels show the colour-magnitude diagram of
NGC 6362\index{NGC 6362} \cite{Piotto2002}, Leo$\,$II\index{Leo$\,$II dwarf galaxy} dwarf galaxy \cite{Held2005},
and Fornax dwarf galaxy\index{Fornax dwarf galaxy} \cite{Holtzman2006d}. The ellipse approximately
traces the BSS region in NGC 6362 and Leo$\,$II, whereas for the Fornax
dwarf contamination by the young stars forbids reliable BSS estimates.\label{momfig:BSS-GC-LeoII-Fornax}}
\end{figure}

Colour-magnitude diagrams\index{colour-magnitude diagram} of typically old dwarf spheroidal galaxies
show the presence of a well-separated blue plume of stars that very
much resembles an old BSS population, as that observed in globular
and open clusters (see the early studies of Mateo et al. \cite{Mateo1991a,Mateo1995c}). However, in the context of dwarf galaxies
one cannot exclude that blue plume stars\index{blue plume} may include genuinely young
main sequence (MS) stars, i.e. a residual star forming activity. The
BSS--young MS ambiguity is simply a hard quest (see discussions in
\cite{Hurley-Keller1998,Aparicio2001c,Carrera2002e}). 

Photometric\index{photometry} techniques like the ($U-B$), ($B-V$) colour-colour diagram\index{colour-colour diagram}
are widely used for Galactic Halo BSS and horizontal branch\index{horizontal branch} (HB) studies. Respectively,
the two colours are proxies for metallicity and temperature and are
useful in separating Halo\index{Halo} BSS from blue horizontal branch stars. This
technique, however, is not necessary for dwarf galaxies studies, because
the dwarf galaxies stellar populations are basically projected at
the same distance from us. Thus the disentangling of the BSS population
in dwarf galaxies relies entirely on the colour-magnitude diagram (as
is the case for Galactic globular and open clusters), and consists
of a selection region within certain luminosity and colour boundaries.

The luminosity function of BSS in globular clusters has been found
to increase from a luminosity cutoff at M$_{V}\sim1.9$ down to M$_{V}\sim4.0$,
at the ancient MS turn-off level \cite{FusiPecci1992} while
the temperature of BSS are between $\sim6000-7500\,$K. An examination
of Fig.$\,$\ref{momfig:BSS-GC-LeoII-Fornax} shows that the colour and
luminosity extensions of the Leo$\,$II blue plume population falls within
the BSS limits in globular clusters. On the other hand, the right
panel of Fig.$\,$\ref{momfig:BSS-GC-LeoII-Fornax} summarises the \emph{BSS--young
stars} ambiguity in dwarf galaxies. The Fornax dwarf galaxy is known
to host a recent star formation episode, that occurred some $200$ Myr
ago \cite{Saviane2000d}, and the diagram from the 
\emph{HST} survey \cite{Holtzman2006d} shows that in such cases one \emph{cannot} properly reach
the ancient MS turn-off level without the inclusion of contaminant
young stars. Thus, the selection of a dwarf galaxies sample for BSS
studies must filter out all those galaxies that \emph{do not} allow
a clear detection of the ancient MS turn-off level. 

In this regards, one should bear in mind that the fainter end of the
BSS sequence extends to $\sim0.6$ magnitude \emph{below} the ancient
MS turn-off level (e.g. the case of M55\index{M55} by \cite{Mandushev1997}).
Given the small number statistics of the BSS stars in dwarf galaxies,
we caution that a conservative BSS selection region (i.e. a bright
M$_{V}\sim3.0$ cutoff that does not reach the ancient main sequence
turn-off level) would heavily under-estimate the BSS frequency in
dwarf galaxies. For example, Mapelli et al. \cite{Mapelli2007b} (using a conservative
BSS selection regions) derive a fraction of HB stars over BSS of $F_{HB}^{BSS}=log(N_{BSS}/N_{HB})\approx-0.65$
for the Draco\index{Draco dwarf galaxy} and Ursa Minor dwarf galaxies\index{Ursa Minor dwarf galaxy}. This is well below the
value derived in \cite{Momany2007} of $F_{HB}^{BSS}\approx+0.1$,
based, however, on deeper and more complete photometric catalogues that
allowed reaching the oldest turn-off level. We note that our 
BSS frequencies \cite{Momany2007} for the two galaxies were recently confirmed by Zhao et al. \cite{Zhao2012bl} (see also the discussion in \cite{Clarkson2011c}). Admittedly however,
there exists no standardised selection process ultimately defining
the BSS selection region for dwarf galaxies, and much is left to the
discretion of the investigators.

\subsection{The BSS Identification in the Galactic Halo\label{momsub:The-BSS-MW-Halo}}

The low-density environment of dwarf galaxies is very similar to that
of the Galactic Halo, hence it is no surprise that a comparison between
their BSS populations is made. In the absence of kinematic and chemical
studies of BSS in dwarf galaxies, the comparison to the Galactic Halo
BSS properties is limited only to the BSS specific frequency. In this
regards, it is worth to spend a few words on the identification of the
Galactic Halo BSS. The chapter by G. Preston (Chap. 4) reports a detailed presentation
of the (i) photometric; (ii) spectro-photometric; and (iii) spectroscopic
criteria used to extract and \emph{disentangle} the BSS from the horizontal
branch population. The spectroscopic survey by {Preston} \& {Sneden} \cite{Preston2000e}
remains a reference point for Milky Way field \emph{blue metal-poor}\index{metal-poor star}
(BSS) studies, and they conclude that over $60\%$ of their sample
is made up by binaries, and that at least $50\%$ of their blue metal-poor
sample are BSS. We all refer to their results for the BSS frequency
in the Galactic Halo: $N_{BSS}/N_{BHB}=4$.

The Halo BSS frequency however has been derived relying on a composite
sample of only $62$ blue metal-poor stars that are: (i) distributed
at different line of sights; (ii) at different distances; and most
importantly, (iii) for which no observational BSS-HB star-by-star
correspondence can be established (see also \cite{Ferraro2006a}). 
Most importantly, one has to bear in mind that the normalised horizontal
branch population is limited to the blue HB \emph{only}, as these
are the only HB component that can be disentangled from the Thin/Thick
Disc population. On the other other hand, the majority of BSS frequency
studies for globular clusters and dwarf galaxies refer to the \emph{entire}
HB population, i.e. including the blue, the variable and the red components
of the HB. Consequently, one has to bear in mind that the adopted
$log(N_{BSS}/N_{BHB})\approx0.6$ value for the Galactic Halo is actually
a \emph{high} upper limit, as indeed the inclusion of red horizontal
branch stars would lower this value. Indeed, a simple check of the
M31 Halo stellar populations, as recently surveyed by the Hubble space
telescope, shows a conspicuous red horizontal branch population (see
Fig.2 in \cite{Brown2008}). Clearly, one has to account for differences
in the star formation and chemical enrichment histories of the two
spiral galaxies, nevertheless the Halo stellar populations of the
two galaxies cannot be very dissimilar. 

\begin{figure}
\begin{centering}
\includegraphics[width=119mm]{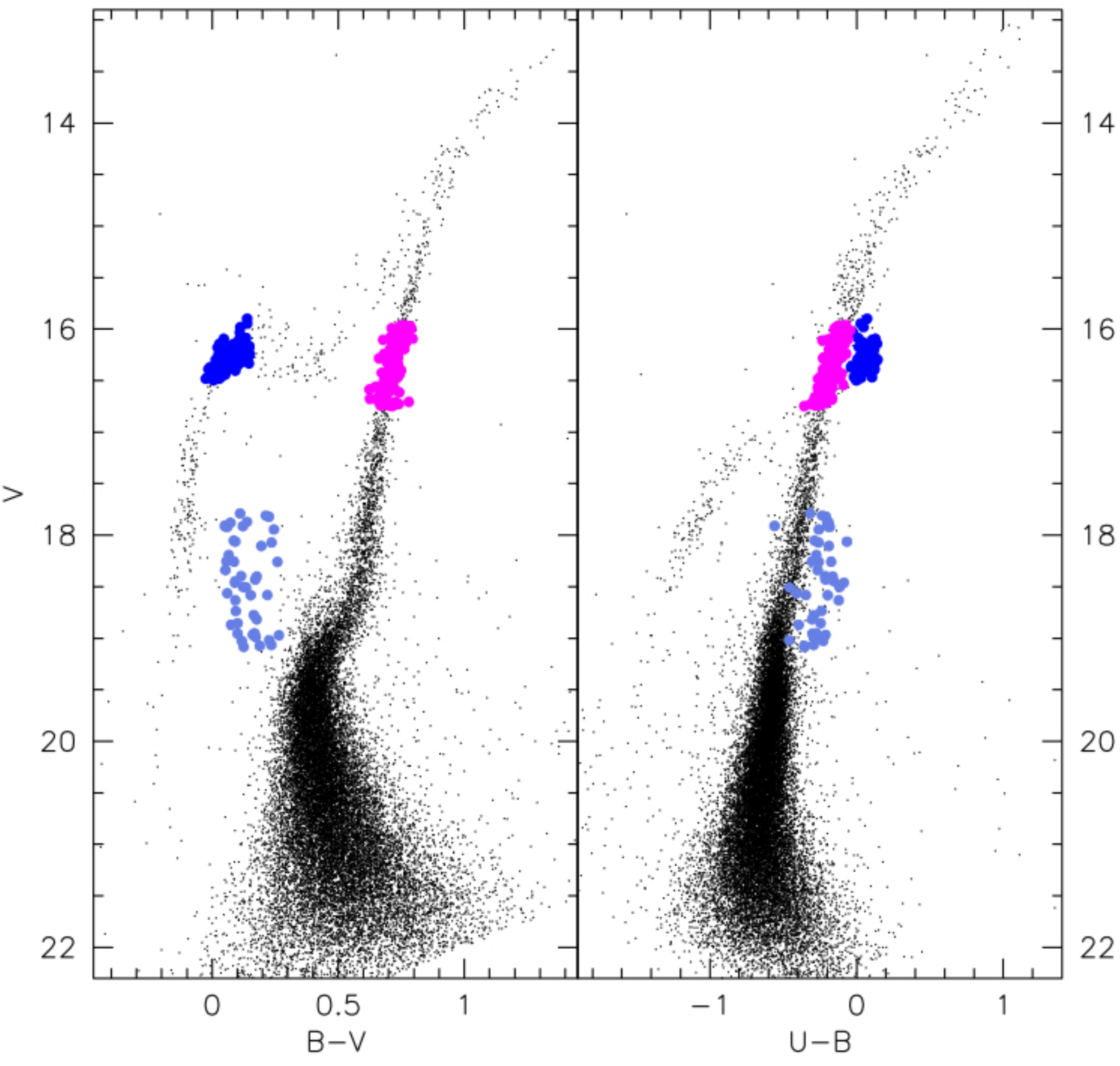}
\par\end{centering}

\caption{\emph{HST} colour-magnitude diagrams of NGC~7078\index{NGC 7078} in the F336W, F439W and F555W
filters, showing a selection of BSS, horizontal and red giant\index{red giant} branch
populations. The right panel shows the same selection in the ($U-B$)
plane, highlighting how the BSS and horizontal branch stars (sharing
almost the same temperature) display the so-called \emph{red-incursion}
\cite{Momany2003b}, which is an instrumental effect.\label{momfig:N7078-leak}}
\end{figure}

Lastly, we point to an often neglected effect when surveying the Galactic
Halo for BSS populations. Figure$\,$\ref{momfig:N7078-leak} displays
the HST colour-magnitude diagram of NGC 7078\index{NGC 7078}. The left panel shows a
typical selection of BSS, HB, and RGB populations and these are later
searched in the $V,$ ($U-B$) plane (right panel). The later diagrams
shows the un-expected and un-physical feature (the so-called \emph{red-incursion,}
see \cite{Momany2003b}) where blue HB stars \emph{can} appear
redder than their red giant equivalents. This feature has been explained
as due to a particular dependence of the adopted $U$ filter (in any
given photometric system) and whether it encompasses the Balmer jump.
Figure$\,$\ref{momfig:N7078-leak} shows that the \emph{red-incursion}
affects also the BSS population, which ends up at redder colours with
respect to red giants at the same luminosity level. The occurrence
and extent of the \emph{red-incursion} shows a dependence of the employed
$U$ and $B$ filters on the star's effective temperature, gravity,
and metallicity. In particular, the \emph{red-incursion} is stronger
(i.e. shows \emph{redder} extent) for lower metallicities stars (see
Fig.5 of \cite{Momany2003b}). Thus, photometric techniques employing
$U,B$ filters for BSS and HB surveys in the Galactic Halo should
take this into account.

\section{BSS Specific Frequency in Dwarf Galaxies}

\subsection{The Dwarf Galaxies Sample}

The large on-sky projection of the closest dwarf galaxies and the
faintness of the BSS in the distant ones preclude the availability
of \emph{a single} \emph{homogeneous and large-area} photometric data-set
addressing the BSS population. Although a significant effort has been
devoted to such purpose (e.g. the HST/WFPC2 archival survey by \cite{Holtzman2006d})
the small field of view of the HST may veil some important spatial
distribution gradients of specific stellar populations. For example
the Leo$\,$I \emph{HST}\index{Hubble Space Telescope} study of \cite{Gallart1999c} did not show the
presence of the blue horizontal (HB) branch population, and this led
to the conclusion that LeoI might have delayed its first epoch of
star formation. However, a \emph{wide-area} coverage by \cite{Held2000}
\emph{did} reveal a conspicuous, ancient, HB population. Thus, in
the context of BSS studies, spatial distribution gradients are important
and should be to accounted for by wide-area surveys.

For example, the largest available catalogue for the Sagittarius dwarf
galaxy\index{Sagittarius dwarf galaxy} is that of Monaco et al. \cite{Monaco2003a}, covering $\sim1^{\circ}$
square degree. However, the Sagittarius dwarf has a core radius of
$\sim3.7^{\circ}$, thus the above-mentioned catalogue \cite{Monaco2003a} covers only
$\sim3.5\%$ of the galaxy, or $\sim6\%$ of its stellar populations.
Hence, any estimate of its BSS frequency is to be taken with caution,
especially that the inner $\sim14^{\prime}\times14^{\prime}$ region
has to be excluded to account for the coincidence of the galaxy's
centre with the position of its globular cluster M54\index{M54}. Moreover, a
delicate aspect of estimating the BSS frequency involves estimating
the Galactic foreground/background contribution in the covered area.
This is particularly important for galaxies in certain line of sights
(e.g. the Sagittarius dwarf suffering severe Galactic Bulge contamination).
To estimate the Galactic contribution in a homogeneous way the \emph{TRILEGAL}\index{TRILEGAL}
code \cite{Girardi2005} was used. This online tool provides synthetic
stellar photometry of the Milky Way components (Disc, Halo, and Bulge\index{Bulge}),
and star counts were performed on the simulated diagrams (using the
same selection boxes) and these subtracted from the observed HB and
BSS star counts for the dwarf galaxies.

There are 12 dwarf galaxies for which a reliable BSS frequency could be determined. The basic properties of the selected galaxies (taken from
\cite{Mateo1998} and \cite{McConnachie2012a}) are summarised
in Table \ref{momtab:The-basic-properties}, and respectively report
the absolute visual magnitude\index{visual magnitude}, the absolute distance modulus\index{distance modulus}, the
reddening\index{reddening}, the central $V$ surface brightness\index{surface brightness}, the core and half-light
radius\index{half-light
radius} and the stellar mass of the galaxy. The BSS specific frequency
--- calculated as $F_{HB}^{BSS}=\log(N_{Bss}/N_{HB})$ --- is reported
in column $\,3$. We emphasise that: (i) photometric incompleteness
corrections; (ii) foreground/background subtraction; (iii) possible
overlap between old and intermediate age stellar population around
the HB level; and (iv) confusion between BSS and normal MS stars,
are \emph{unavoidable} sources of error that affect any analysis addressing
the BSS frequency in dwarf galaxies. The reported error bars (reported
in column $4$) account for the propagation of the Poisson errors
on the star counts, but mostly reflect the dependence on the uncertainty
in properly defining the HB and BSS selection boxes. In particular,
the reported BSS specific frequency have been either taken directly
from the literature (as is the case for the Cetus\index{Cetus dwarf galaxy} and Tucana dwarf
galaxies\index{Tucana dwarf galaxy}, \cite{Monelli2012}) or from estimates based on photometric
catalogues made available for the following objects: (i) Sextans\index{Sextans dwarf galaxy} \cite{Lee2003u};
(ii) Ursa Minor\index{Ursa Minor dwarf galaxy} \cite{Carrera2002e}; (iii) Sculptor\index{Sculptor dwarf galaxy} \cite{Rizzi2003};
(iv) Ursa Major\index{Ursa Major  dwarf galaxy} by \cite{Willman2005a}; (v) Bootes\index{Bootes} \cite{Belokurov2006c};
(vi) Sagittarius\index{Sagittarius dwarf galaxy} by \cite{Monaco2003a}; (vii) Leo$\,$II\index{Leo$\,$II dwarf galaxy} \cite{Held2005};
(viii) Draco\index{Draco dwarf galaxy} \cite{Aparicio2001c}; and (ix) Leo$\,$IV\index{Leo$\,$IV  dwarf galaxy} and Hercules\index{Hercules  dwarf galaxy}
by Milone (\emph{priv. comm.}, based on \emph{HST}\index{Hubble Space Telescope} diagrams).

\begin{figure}
\begin{centering}
\includegraphics[width=119mm]{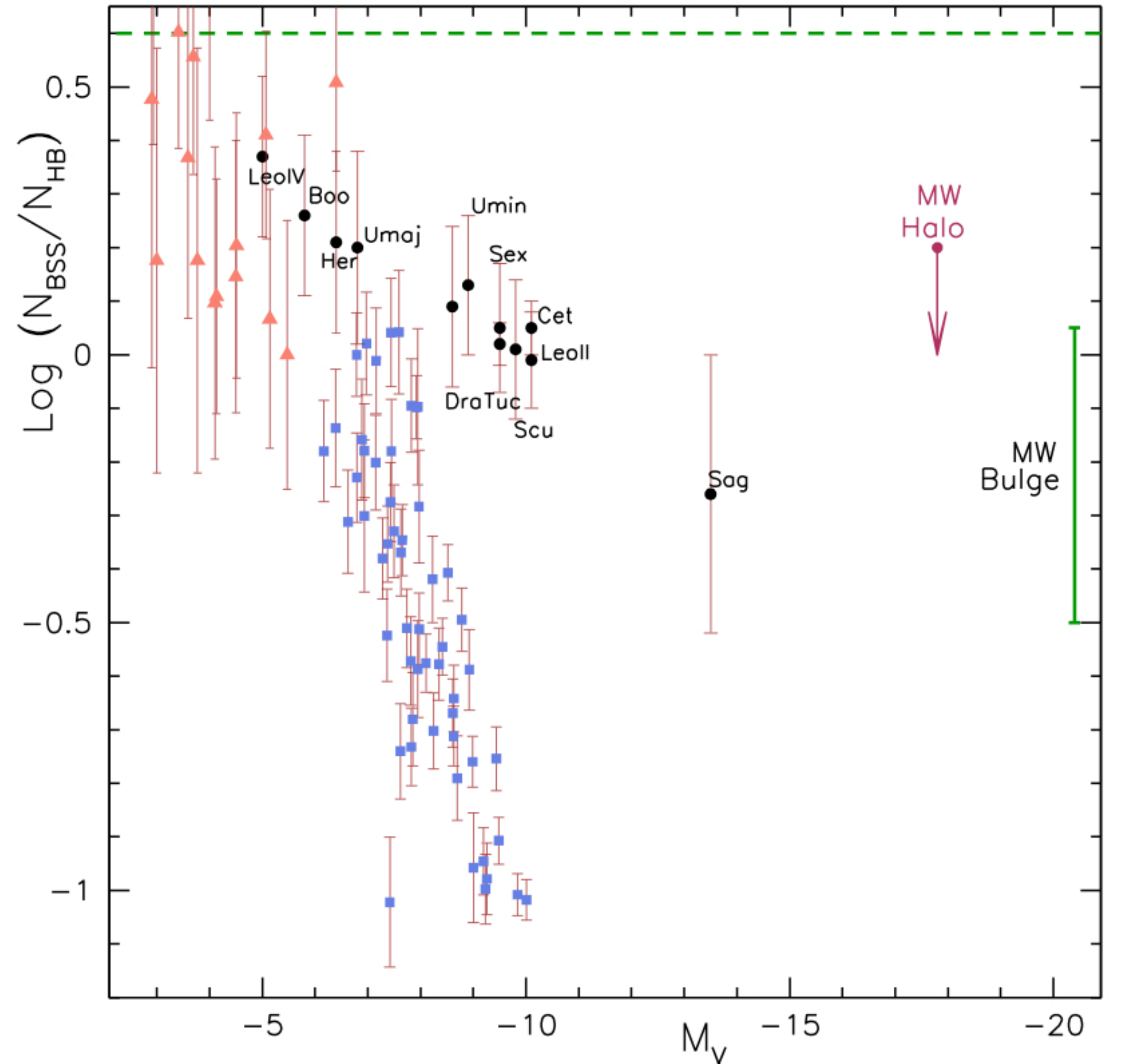}
\par\end{centering}

\caption{The BSS frequency ($F_{HB}^{BSS}$) \emph{vs} M$_{V}$ diagram for
globular clusters \cite{Piotto2004c}, open clusters \cite{2006A&A...459..489D},
and dwarf galaxies \cite{Momany2007}. The horizontal line shows
the mean BSS frequency as derived for the Milky Way field stars by
\cite{Preston2000e}. The range of the BSS frequency for the Milky
Way Bulge \cite{Clarkson2011c} is shown as a bar at M$_{V}=-20.4$.
A preliminary \emph{upper limit} for the Milky Way Halo BSS frequency
(as normalised to \emph{only} the blue horizontal branch stars) is
shown at M$_{V}=-17.8$. \label{momfig:Freq-vs-Mv}}
\end{figure}

\begin{table}
\caption{Basic properties of dwarf galaxies for which the BSS frequency
could be reliably derived}
\label{momtab:The-basic-properties}
\begin{tabular}{llccclclc}
\hline\noalign{\smallskip}
{\scriptsize Name}  & {\scriptsize $M_{V}$}  & {\footnotesize $Log(F_{HB}^{BSS})$}  & {\tiny $\sigma_{Log(F_{HB}^{BSS})}$ } & {\scriptsize ($m-M$)$_{\circ}$ }  & {\scriptsize $E_{(B-V)}$ }  & {\scriptsize $\mu_{V}${[}mag/$\square$$^{\prime\prime}${]} }  & {\scriptsize $r_{h}${[}$^{\prime}${]}}  & {\scriptsize M$_{\star}${[}$10^{6}$M$_{\odot}${]}}\tabularnewline
\noalign{\smallskip}\svhline\noalign{\smallskip}
{\scriptsize Bootes}  & {\scriptsize $-$5.8}  & {\scriptsize 0.26 } & {\scriptsize 0.15}  & {\scriptsize 18.9}  & {\scriptsize 0.000}  & {\scriptsize 28.05}  & {\scriptsize 12.6}  & {\scriptsize 0.029}\tabularnewline
{\scriptsize Ursa Major }  & {\scriptsize $-$6.8 }  & {\scriptsize 0.20} & {\scriptsize 0.18 }  & {\scriptsize 17.1 }  & {\scriptsize 0.150 }  & {\scriptsize 27.80 }  & {\scriptsize 11.3 }  & {\scriptsize 0.014}\tabularnewline
{\scriptsize Draco }  & {\scriptsize $-$8.6 }  & {\scriptsize 0.09}  & {\scriptsize 0.15 }  & {\scriptsize 19.5 }  & {\scriptsize 0.030 }  & {\scriptsize 25.30 }  & {\scriptsize 10.0 }  & {\scriptsize 0.290}\tabularnewline
{\scriptsize Ursa Minor }  & {\scriptsize $-$8.9 }  & {\scriptsize 0.13}  & {\scriptsize 0.13 }  & {\scriptsize 19.0 }  & {\scriptsize 0.030 }  & {\scriptsize 25.50 }  & {\scriptsize 8.2 }  & {\scriptsize 0.290}\tabularnewline
{\scriptsize Sextans }  & {\scriptsize $-$9.5 }  & {\scriptsize 0.05}  & {\scriptsize 0.12 $ $}  & {\scriptsize 19.7 }  & {\scriptsize 0.040 }  & {\scriptsize 26.20 }  & {\scriptsize 27.8 }  & {\scriptsize 0.440}\tabularnewline
{\scriptsize Sculptor }  & {\scriptsize $-$9.8 }  & {\scriptsize 0.01}  & {\scriptsize 0.13 }  & {\scriptsize 19.7 }  & {\scriptsize 0.000 }  & {\scriptsize 23.70 }  & {\scriptsize 11.3 }  & {\scriptsize 0.230}\tabularnewline
{\scriptsize LeoII }  & {\scriptsize $-$10.1 }  & {\scriptsize $-$0.01 } & {\scriptsize 0.09 }  & {\scriptsize 21.6 }  & {\scriptsize 0.030 }  & {\scriptsize 24.00 }  & {\scriptsize 2.6 }  & {\scriptsize 0.740}\tabularnewline
{\scriptsize Sagittarius }  & {\scriptsize $-$13.5 }  & {\scriptsize $-$0.26}  & {\scriptsize 0.26 }  & {\scriptsize 17.1 }  & {\scriptsize 0.150 }  & {\scriptsize 25.40 }  & {\scriptsize 342.0 }  & {\scriptsize 21.00}\tabularnewline
{\scriptsize Cetus}  & {\scriptsize $-$10.1 }  & {\scriptsize 0.05}  & {\scriptsize 0.05 }  & {\scriptsize 24.5 }  & {\scriptsize 0.030 }  & {\scriptsize 25.00 }  & {\scriptsize 3.2 }  & {\scriptsize 2.600}\tabularnewline
{\scriptsize Tucana }  & {\scriptsize $-$9.5 }  & {\scriptsize 0.02}  & {\scriptsize 0.04 }  & {\scriptsize 24.7 }  & {\scriptsize 0.030 }  & {\scriptsize 25.00 }  & {\scriptsize 1.1 }  & {\scriptsize 0.560}\tabularnewline
{\scriptsize Leo IV }  & {\scriptsize $-$5.0 }  & {\scriptsize 0.49}  & {\scriptsize 0.16 }  & {\scriptsize 20.9 }  & {\scriptsize 0.026 }  & {\scriptsize 27.50 }  & {\scriptsize 4.6 }  & {\scriptsize 0.019}\tabularnewline
{\scriptsize Hercules}  & {\scriptsize $-$6.4}  & {\scriptsize 0.21} & {\scriptsize 0.17}  & {\scriptsize 20.6}  & {\scriptsize 0.063 }  & {\scriptsize 27.20}  & {\scriptsize 8.6 }  & {\scriptsize 0.037}\tabularnewline
\noalign{\smallskip}\hline\noalign{\smallskip}
\end{tabular}
\end{table}

\subsection{The F$_{HB}^{BSS}-$M$_{V}$ Anti-correlation}

For a wider perspective on the BSS specific frequency in dwarf galaxies,
a comparison is made with that derived for Galactic globular\index{globular cluster} and open
clusters\index{open cluster}. Of the original compilation of BSS in Galactic open cluster
by de Marchi et al. \cite{2006A&A...459..489D} we filter out clusters for which less
than two BSS stars were found. On the other hand, the original compilation
of $\sim3000$ BSS in $56$ Galactic globular clusters by Piotto et al. \cite{Piotto2004c}
was complemented by BSS frequencies for three interesting additional clusters,
namely: (i) NGC 1841\index{NGC 1841} \cite{Saviane2003}, which is the Large
Magellanic Cloud\index{Large Magellanic Cloud} (LMC) most metal-poor and most distant (10 kpc from
the LMC bar) cluster; (ii) NGC 2419\index{NGC 2419}, which is a massive Milky Way
cluster at 90 kpc from the Galactic centre, suspected to be a dwarf
galaxy; and (iii) $\omega$~Cen\index{$\omega$ Centauri}, which is the most enigmatic Milky
Way cluster, also suspected to be an extra-Galactic dwarf galaxy.
The three data-points are based on deep \emph{HST}/ACS\index{Hubble Space Telescope}, WFPC2 and ACS archival
data, respectively. Focusing our attention only on the BSS frequency
of Galactic globular cluster, in particular for the three additional clusters,
Fig.$\,$\ref{momfig:Freq-vs-Mv} shows that the so-called $F{}_{HB}^{BSS}-$M$_{V}$
anti-correlation \cite{Piotto2004c,Davies2004a}
is basically \emph{universal} for all globular clusters, regardless
of their specific complexity and origin. The observed anti-correlation
implies that more massive globular clusters are surprisingly BSS-deficient,
as if their high collision\index{collision} rate had no correlation with the production
of collisional binaries. In particular, the $F{}_{HB}^{BSS}-$M$_{V}$
anti-correlation was explained by {Davies}, {Piotto} \& {de Angeli} \cite{Davies2004a} (see also \cite{Mapelli2004,Mapelli2006c}) in the following manner: the number of
BSS produced via collisions tends to increase with cluster mass, becoming
the dominant formation channel for clusters with M$_{V}\leq-8.8$.
On the other hand, the BSS number originating from primordial binaries
should decrease with increasing cluster mass. Accounting for these
two opposite trends and binary evolution, the models by Davies et al. \cite{Davies2004a}
reproduce the observed BSS population, whose total number seems independent
of the cluster mass.

When plotting the BSS frequency in dwarf galaxies, Fig.$\,$\ref{momfig:Freq-vs-Mv}
shows the following general trends: (i) dwarf galaxies with M$_{V}\leq-8.0$
possess a relatively higher BSS frequency with respect to globular
clusters with similar luminosities; and (ii) the lowest luminosity
dwarf galaxies with $-8.0\leq M_{V}\leq-5.0$ show BSS frequencies
that are fully compatible with that observed in open clusters. This
compatibility between dwarf galaxies and open clusters may suggest
that there exists a ``saturation\textquotedblright{} in the BSS frequency
(at \emph{$F_{HB}^{BSS}\approx0.3-0.4$}) for the lowest luminosity
systems (both for open clusters and dwarf galaxies). In this regards,
we note that the globular clusters distribution shows an abrupt cut
at M$_{V}\sim-6.0$. This is only a selection effect, and we remind
the reader that there are a dozen of clusters with M$_{V}\gtrsim-6.0$
that were not included in the Piotto et al. \cite{Piotto2004c} survey. It is of
great importance to fill this gap and test the hypothesis that $all$
stellar systems show a saturation of the BSS frequency at \emph{$F_{HB}^{BSS}\approx0.3-0.4$}.

There is a hint of such \emph{universal upper limit} in the study
by Sollima et al. \cite{Sollima2008c} who derive high BSS frequencies for three
clusters with M$_{V}\approx-5.0$. Similarly, Santana et al. \cite{Santana2012}
derive high BSS frequencies for the faintest tail of the globular
clusters at M$_{V}\approx-5.0$. Unfortunately, these clusters could
not be added to Fig.$\,$\ref{momfig:Freq-vs-Mv} because Sollima et al. \cite{Sollima2008c}
normalised their BSS star counts with respect to the main sequence
stars, while the study of Santana et al. \cite{Santana2012} used the red giant branch
stars.

Overall, the dwarf galaxies sample shows a hint of a proper $F_{HB}^{BSS}-$M$_{V}$
anti-correlation. In \cite{Momany2007}, and relying on a smaller
sample of $8$ galaxies, the statistical significance of a $F{}_{HB}^{BSS}-$M$_{V}$
anti-correlation was explored, and the probability that a random sample
of uncorrelated experimental data points would have yielded a linear-correlation
coefficient of $0.984$ was found to be extremely low ($\leq10^{-6}$).
With respect to the above \cite{Momany2007} study, the present dwarf galaxy
sample has four new solid entries, that populate the extreme ends
of the dwarf galaxies luminosity distribution. The BSS frequency for
the Tucana and Cetus dwarfs at around M$_{V}\approx-10.0$ are taken
directly from Monelli et al. \cite{Monelli2012}, whereas for the Leo${\,}$IV and Hercules
at around M$_{V}=-5.0$, the frequencies were derived. Repeating the
same exercise, the statistical significance of the hinted $F{}_{HB}^{BSS}-$M$_{V}$
anti-correlation still holds. Nevertheless, it is populating the dwarf
galaxy sample with the ultra-faint dwarfs with M$_{V}\approx-4.0$
that the anti-correlation can be firmly established. In this regards,
we note the extreme difficulty in the process of identification of
BSS samples in such galaxies, as this heavily relies on the quality
of photometric catalogues. For example, the BSS frequency for the Leo$\,$IV
and Hercules dwarf galaxies was derived only thanks to yet unpublished
\emph{Hubble Space Telescope}\index{Hubble Space Telescope} deep diagrams that reveal with confidence the
presence of BSS in these systems. On the other hand, ground-based
observations of the same two galaxies did not allow a reliable estimate
of the BSS frequency. 

Lastly, one should also keep in mind that for the ultra-faint dwarf
galaxies even the absolute luminosity of the system is subject to
significant variations, i.e. the inclusion or not of few red giant
stars (whose census become very sensitive to foreground contamination)
has reflected on changes of the order of $\sim0.6$ magnitude for
some galaxies. 

\textbf{\emph{The specific case of the Carina dwarf galaxy}}: previously
we emphasised the obvious need to exclude the group of gas-rich, star
forming dwarf galaxies from our sample. To this constraint, one can
also add the group of dwarf galaxies with known, dominant, intermediate-age
population. Indeed, the presence of intermediate age population of
around $\sim5$ Gyr would inevitably interfere with the BSS selection
process and impose the shifting of the BSS box to brighter magnitudes.
A perfect example of such a case is the Carina dwarf galaxy\index{Carina dwarf galaxy}. The colour-magnitude
diagrams\index{colour-magnitude diagram} of {Hurley-Keller}, {Mateo} \& {Nemec} \cite{Hurley-Keller1998} and Bono et al. \cite{Bono2010e} show
clearly the presence of multiple subgiant\index{subgiant} branches separated by $\sim0.5-0.8$
magnitude. The Carina star formation history reconstructed by Rizzi et al. \cite{Rizzi2003e}
shows that the bulk of the star formation has taken place in an episode
at around $\sim6$ Gyr. Deriving a BSS frequency for Carina (as done
for the other galaxies but with the limitation of not reaching the
faintest turn-off level) results in a \emph{lower} limit of $F{}_{HB}^{BSS}=0.4$,
which at M$_{V}=-9.1$ would \emph{still} reflect a high BSS frequency
with respect to other dwarf galaxies of similar luminosity (e.g. the
Tucana dwarf at M$_{V}=-9.5$ has $F{}_{HB}^{BSS}=0.0$). This has
triggered the use of the $F{}_{HB}^{BSS}-$M$_{V}$ anti-correlation
as a diagnostic for the presence of young stellar populations in other
galaxies. For example, examining the diagrams of the Canes Venatici$\,$I
dwarf\index{Canes Venatici$\,$I
dwarf galaxy} (M$_{V}=-8.6$), Martin et al. \cite{Martin2008} estimate the BSS/blue plume
frequency to be $F{}_{HB}^{BSS}=0.5$, and along with arguments concerning
the spatial distribution of this population they conclude that it
is best understood as a young stellar population.

\textbf{\emph{The specific case of the Milky Way Bulge\index{Bulge}}}\emph{: }thanks
to recent and multi-epoch \emph{Hubble Space Telescope} observations of the
Galactic Bulge, Clarkson et al. \cite{Clarkson2011c} beautifully managed to proper-motion
decontaminate the Bulge stellar populations from the foreground disc
contribution. Their goal was to investigate the presence (or not)
of a genuine young stellar population in the Galactic Bulge. The cleaned
colour-magnitude diagram however resembles that of a typical old stellar
population with a scarce population of seemingly BSS. They \cite{Clarkson2011c}
estimate the BSS specific frequency and (as done in this and many
studies) use the convention of normalising the BSS numbers as a function
of the entire horizontal branch population. Assuming a M$_{V}=-20.4$
for the Bulge they conclude that the limits of their BSS frequency
is consistent (see Fig.$\,$\ref{momfig:Freq-vs-Mv}) with the general
trend displayed by the $F{}_{HB}^{BSS}-$M$_{V}$ anti-correlation,
suggested in our earlier work \cite{Momany2007}. The addition of the Bulge extends
the $F_{HB}^{BSS}-$M$_{V}$ anti-correlation by over $7$ magnitudes
to the extreme massive systems. As emphasised by Clarkson et al. \cite{Clarkson2011c},
the interpretation of this agreement remains unclear and awaits further
confirmation of the anti-correlation itself. 

Lastly, availing recent \emph{SEGUE\index{SEGUE}} survey results for BSS and blue-HB candidates
(Santucci R., priv. comm.), a rough and preliminary BSS frequency
for the Galactic Halo of $F_{BHB}^{BSS}\sim0.2$ was derived involving
$9,000$ and $5,600$ stars. To assign an absolute luminosity of the
Galactic Halo\index{Halo}, we roughly assume that the family of Galactic globular
clusters contribute by $1\%$ of the total luminosity, and derive
M$_{V}\sim-17.8$. As argued in Sec.\ref{momsub:The-BSS-MW-Halo}, normalising
the BSS star counts to only blue-HB stars would provide a strong upper
limit. Nonetheless, we add this point to Fig.$\,$\ref{momfig:Freq-vs-Mv}
for comparison purposes and speculate that the Galactic Halo BSS frequency
might approach values of $F_{HB}^{BSS}\sim0.0$ and lower. All together,
the addition of 4 new dwarf galaxies, along with that of the Galactic
Bulge and evidence from the Anomalous Cepheid frequency\index{Anomalous Cepheid variable} (see Sec.\ref{momsec:The-Progeny-of-BSS})
seem to confirm the $F{}_{HB}^{BSS}-$M$_{V}$ anti-correlation.

\begin{figure}
\begin{centering}
\includegraphics[width=119mm]{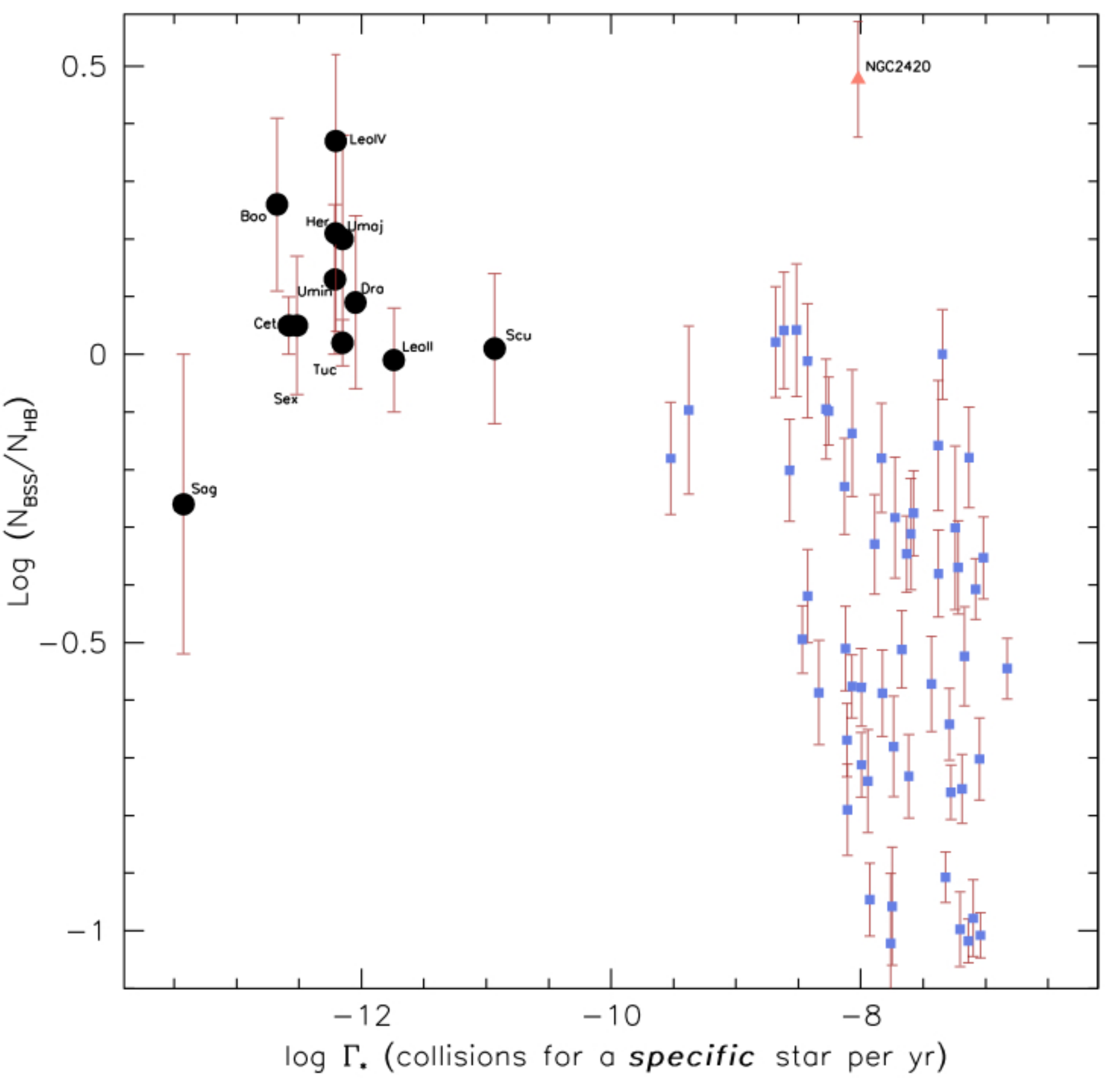}
\par\end{centering}

\caption{The BSS frequency as a function of the stellar collision factor for
a specific star per year. Globular clusters are plotted as filled
squares, while dwarf galaxies are plotted as filled circles. The filled
triangle is that of an open cluster. \label{momfig:collision-factor vs BSS-freq}}
\end{figure}

\subsection{The Significance of the Anti-correlation}

The blue plume\index{blue plume} of dwarf galaxies (currently not experimenting star
formation episodes) does not allow clear cut indications on whether
it represents: (i) a genuine BSS population (as that observed in globular
clusters); (ii) a young population of recently ($\leq2$ Gyr) formed
stars; or (iii) a combination of both. To tackle this ambiguous problem,
and given the faintness of the blue plume stars, one could only address
the number frequency of this population and search for general trends
as a function of the parent galaxy parameters. To attempt such analysis,
one \emph{must} tacitly assume that the blue plume is made of a genuine
BSS population (perform the BSS star counts and normalisation) and
then search for a correlation, say, with the parent galaxy luminosity
(a proxy of mass). Should the assumption be incorrect --- and the blue
plume population is instead made of (or contaminated by) young MS
stars --- then one would expect to find a \emph{correlation} between
the BSS frequency and the galaxy luminosity (i.e. more massive galaxies
tend to preserve a low level of residual star formation rate and hence
possess a larger fraction of young stars). On the other hand, a flat
BSS frequency distribution would need an ad hoc scenario where more
massive galaxies \emph{conspire} and coordinate their star formation
rate in a way to mimic a rather flat ``BSS'' frequency. Instead,
and with all necessary caution, we find hints of an \emph{anti-correlation}.
This is unexpected and points to a genuine BSS origin of the blue
plume in the studied galaxies. The fact that globular clusters do
show a similar anti-correlation makes it easier to suggest that whatever
mechanism is at work in globular clusters \emph{might} be responsible
for the milder anti-correlation seen in the dwarf galaxies. Granted
the above, does the scenario envisaged by Davies et al. \cite{Davies2004a} apply
also for dwarf galaxies?

The above question implies answering ``\emph{do dwarf galaxies harbour
a significant population of collisional\index{collision} binaries at all}?'' The
answer is \emph{no}. This lies in the intrinsic properties of dwarf
galaxies\index{dwarf galaxy} and their differences from globular clusters\index{globular cluster}. Indeed, it
is enough to recall that the central luminosity density of a dwarf
galaxy (e.g. Ursa Minor\index{Ursa Minor dwarf galaxy}: $0.006\, L_{\odot}$ pc$^{-3}$ at M$_{V}=-8.9$)
is several orders of magnitudes lower than that found in a typical
globular cluster (e.g. NGC 7089\index{NGC 7089}: $\sim8000\, L_{\odot}$ pc$^{-3}$
at M$_{V}=-9.0$). This implies that the collisional parameter of
dwarf galaxies is very low, and unambiguously point to the much slower
dynamical evolution of dwarf galaxies. To further emphasise this last
point, in Figure \ref{momfig:collision-factor vs BSS-freq} we show $F_{HB}^{BSS}$
as a function of a \emph{calculated} quantity: the stellar specific
collision parameter ($log$$\Gamma_{\star}$: the number of collisions
per specific star per year). More specifically, following Piotto et al. \cite{Piotto2004c},
we estimate $log$$\Gamma_{\star}$ from the systems's central surface
density and core size. To these, we could add the $log$$\Gamma_{\star}$
corresponding value of an open cluster, thanks to parameters provided
by Giovanni Carraro (priv. comm.).

The mean collisional parameter of the 12 studied galaxies is $-11.5$.
The lowest value is that for the Sagittarius dwarf, and this is probably
due to its very extended galaxy core. Compared with the mean value
of $\log \Gamma_{\star}=-7.5$ for the globular clusters sample,
the estimated number of collisions per specific star per year in
a dwarf galaxy is $10^{-5}$ times lower. This almost \emph{precludes}
the occurrence of collisional binaries in dwarf galaxies, and one
may conclude that genuine BSS sequences in dwarf galaxies are mainly
made of primordial binaries. Moreover, the overall $\log \Gamma_{\star}$
distribution of the dwarf galaxies shows no obvious correlation with
the BSS frequency. This is in good agreement with the observational
fact (see \cite{Piotto2004c}) that the collisional parameter of
globular clusters also do not show any correlation with the BSS frequency.

The flatter slope of the potential anti-correlation in dwarf galaxies
may be understood in terms of the almost lack of collisional--BSS:
i.e. neither created nor destroyed. Moreover, not all primordial binaries,
now present in a dwarf galaxy, turn into or are already in the form
of BSS. In particular, it is the low exchange encounter probabilities
in environments like the Galactic Halo or dwarf galaxies that guarantees
a friendly environment and a slower consumption/evolution of primordial
binary systems. The BSS production (via evolution off the MS of the
primary and the consequent mass transfer\index{mass transfer} to the secondary that may
become a BSS) is still taking place in the present epoch and this
can explain the high frequency of primordial BSS in dwarf galaxies
as well as the Galactic Halo.

\section{The BSS Radial Distribution and Luminosity Function in Dwarf Galaxies}

As argued in the previous section, the collisional rate of dwarf galaxies
is $10^{-5}$ times lower than in globular clusters. This practically
precludes the BSS \emph{collisional} formation channel in dwarf galaxies,
thereby supporting solely the \emph{mass-transfer} binaries channel.
This conclusion can be confirmed or refuted by examining the radial
distribution and luminosity functions of the BSS population in dwarf
galaxies. Indeed, models of the globular clusters BSS that are \emph{thought}
to originate via the \emph{collisional} channel foresee specific radial
distribution and luminosity functions signature, and such signatures
can be verified in dwarf galaxies.

\begin{figure}
\begin{centering}
\includegraphics[width=119mm]{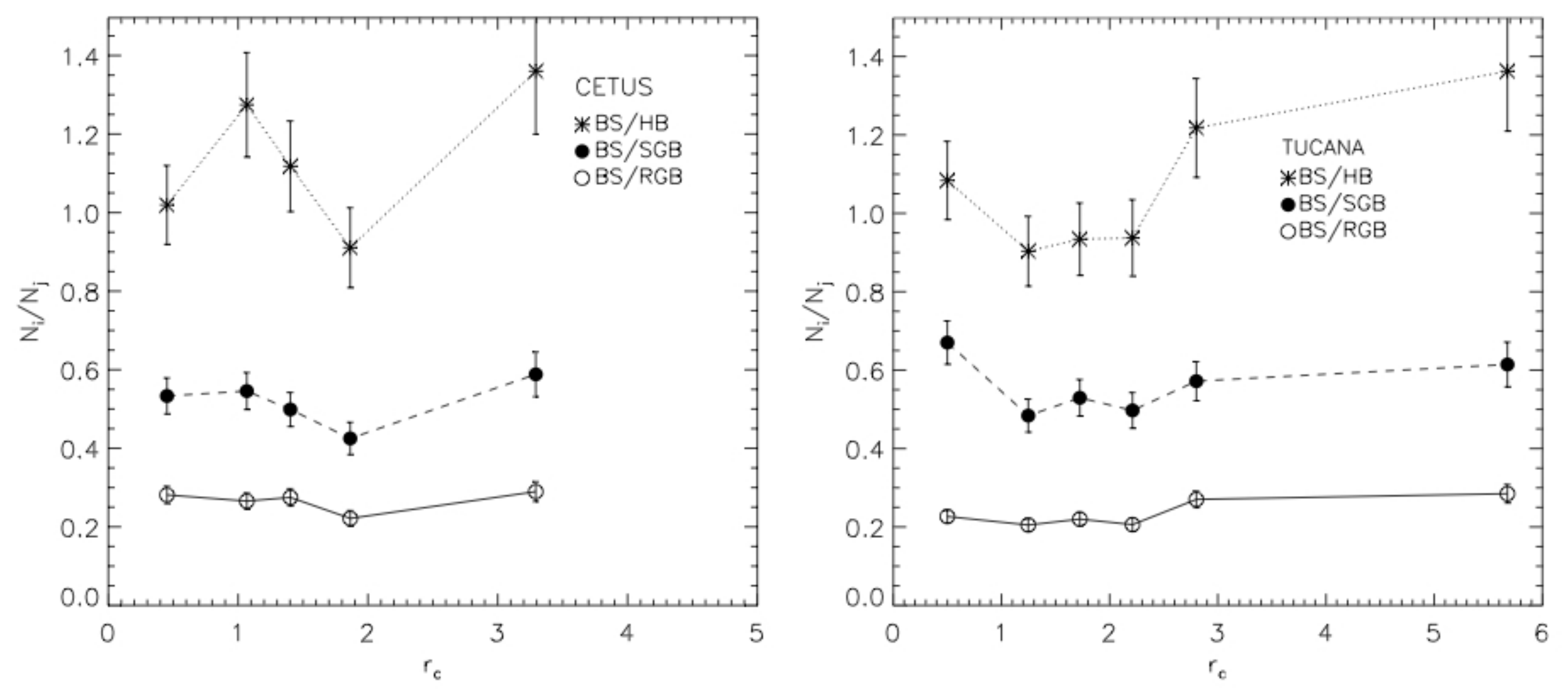}
\par\end{centering}

\caption{The BSS radial distribution as a function of the core galacto-centric
radius for the Cetus\index{Cetus dwarf galaxy} (left) and Tucana\index{Tucana dwarf galaxy} (right) dwarf galaxies. The
absence of a central peak points to the lack of collisional BSS in
dwarf galaxies. This figure is reproduced from \cite{Monelli2012}
with permission by the AAS. \label{momfig:Radial-Distrib-Cetus-Tucana}}
\end{figure}

\subsection{Radial Distribution}

Thanks to their high density cores, globular clusters are the ideal
environment where collisional BSS can, and \emph{must}, form. Interestingly,
the BSS radial distribution of the majority of globular clusters is
bimodal \cite{FusiPecci1992,Ferraro1997e,Zaggia1997,Lanzoni2007,Beccari2008}: showing a central
peak, followed by a gradual decrease until reaching a minimum at intermediate
radii, and then showing a rise at the clusters periphery. This bimodality
was explained \cite{Mapelli2004,Mapelli2006c}
by the joint contribution of: (i) \emph{collisional} binaries, naturally
produced in the highest density regions of the parent cluster, thereby
responsible for the central peak; and (ii) \emph{mass-transfer} binaries
which, left peacefully to evolve in the cluster outskirts, avoid sinking
towards the cluster centre, and produce the outer peak. A minority
of globular clusters \cite{Lanzoni2007a,Beccari2011}
do not show the external rise and are expected to be poor in \emph{mass-transfer}\index{mass transfer}
BSS. Overall, $all$ globular clusters show a central peak in their
BSS radial distribution and this is attributed to the production of
\emph{collisional} BSS\index{collision}. 

The wide-field imaging\index{wide-field imaging} study of the Draco\index{Draco dwarf galaxy} and
Ursa Minor\index{Ursa Minor dwarf galaxy} dwarf galaxies by Mapelli et al. \cite{Mapelli2007b}  shows clearly that the radial distribution
of the BSS population is \emph{flat} (see also \cite{Carrera2002e}),
and hardly consistent with a central peak like that observed in globular
clusters. If ever, the BSS populations in this study \cite{Mapelli2007b}
show a hint of a decrease in the central regions. Mapelli et al. \cite{Mapelli2007b}
conclude that the BSS in the Draco and Ursa Minor dwarf galaxies have
radial distributions that are consistent with the expectations of
their models of
\emph{mass-transfer}\index{mass transfer} BSS \cite{Mapelli2004,Mapelli2006c} . Similar results were obtained for the case
of the Sculptor\index{Sculptor dwarf galaxy} dwarf by \cite{Mapelli2009d}, and Cetus\index{Cetus dwarf galaxy} and Tucana\index{Tucana dwarf galaxy}
by \cite{Monelli2012}. Figure$\,$\ref{momfig:Radial-Distrib-Cetus-Tucana}
shows an excellent example of flat BSS distributions in the Cetus and
Tucana dwarf galaxies (from \cite{Monelli2012}).

\begin{figure}
\begin{centering}
\includegraphics[width=119mm]{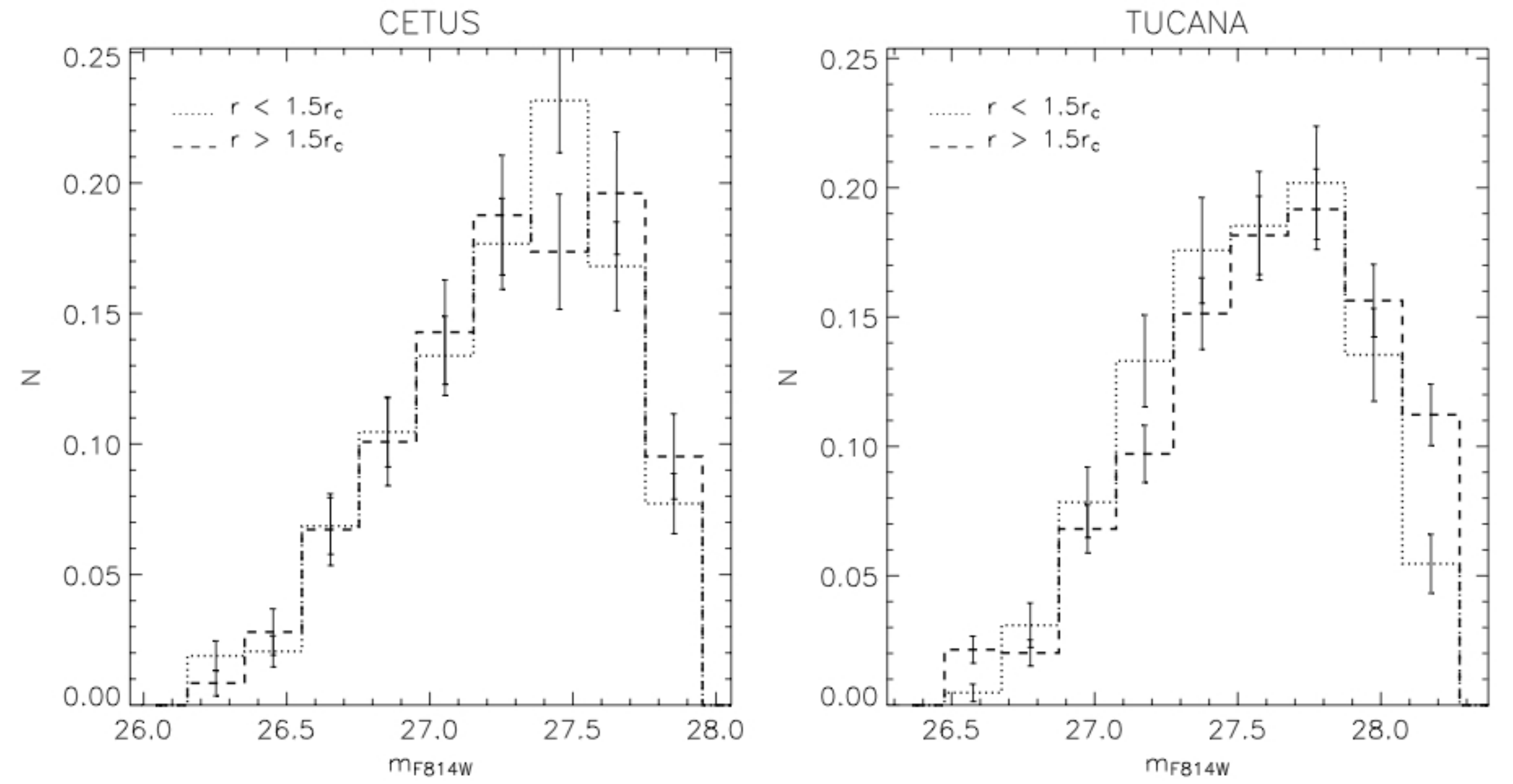}
\par\end{centering}

\caption{The BSS normalised luminosity functions in the Cetus and Tucana dwarf
galaxies. The different lines refer to the BSS stars within (dotted)
and outside (dashed) $1.5\times r_{c}$ from the galaxies centres.
The agreement between the two luminosity functions (for both galaxies)
hints to the absence of collisional--BSS in dwarf galaxies. This figure
is reproduced from \cite{Monelli2012} with permission by the
AAS. \label{momfig:Lum-Function-Cetus-Tucana}}
\end{figure}

\subsection{Luminosity Function}

Monkman et al. \cite{Monkman2006} find that \emph{bright} BSS stars in 47~Tuc\index{47 Tucanae} tend
to be more centrally concentrated. The central distribution of these
bright BSS stars strengthen the idea of a collisional origin. Indeed,
the collisional formation channel should allow the final product (the
BSS star) to retain a large fraction of the original masses involved
in the collisions, hence the brighter luminosities. In terms of the
luminosity function, this translates into a correlation between the
BSS luminosities and their radial distribution. Should the dwarf galaxy
lack the presence of collisional binaries (as we have argued) one
expects no correlation between an inner and outer BSS luminosity functions.

This hypothesis was tested by Mapelli et al. for the Draco,
Ursa Minor dwarf galaxies and for Sculptor \cite{Mapelli2007b,Mapelli2009d},
and by Monelli et al. \cite{Monelli2012} for the Cetus and Tucana dwarfs. With
the only exception of Sextans\index{Sextans dwarf galaxy} \cite{Lee2003u}, all these studies
proved the absence of a correlation between the BSS luminosity function
and the radial distribution. This is best illustrated in Fig.\ref{momfig:Lum-Function-Cetus-Tucana}
taken from \cite{Monelli2012}. For each galaxy, two BSS luminosity
functions were derived and compared. The excellent agreement between
the inner and outer $1.5\times r_{c}$ selections points to the absence
of collisional binaries products, thereby confirming the mass-transfer\index{mass transfer}
binaries as the sole origin of BSS in dwarf galaxies.

\section{Variable BSS in Dwarf Galaxies}

SX Phoenicis\index{SX Phe star} (SX Phe) are a class of Population$\:$II pulsating variables
that exhibit short period ($P<0.1$ days) variations, having spectral
types between A2--F5. SX Phe variables are particularly interesting
because the region in the colour-magnitude diagram where they cross
the instability strip coincides with the BSS location for globular
clusters. Indeed, all SX Phe variables in globular clusters are BSS
stars. SX Phe are, however, $\sim1-2.5$ magnitudes fainter than RR
Lyrae\index{RR Lyrae star} and therefore, in the context of distant dwarf galaxies studies,
are harder to detect. {Mateo}, {Fischer} \& {Krzeminski} \cite{Mateo1998d} and Poretti et al. \cite{Poretti1999}
were the first to report on SX Phe in the Carina dwarf galaxy\index{Carina dwarf galaxy}.

\begin{figure}
\begin{centering}
\includegraphics[width=119mm]{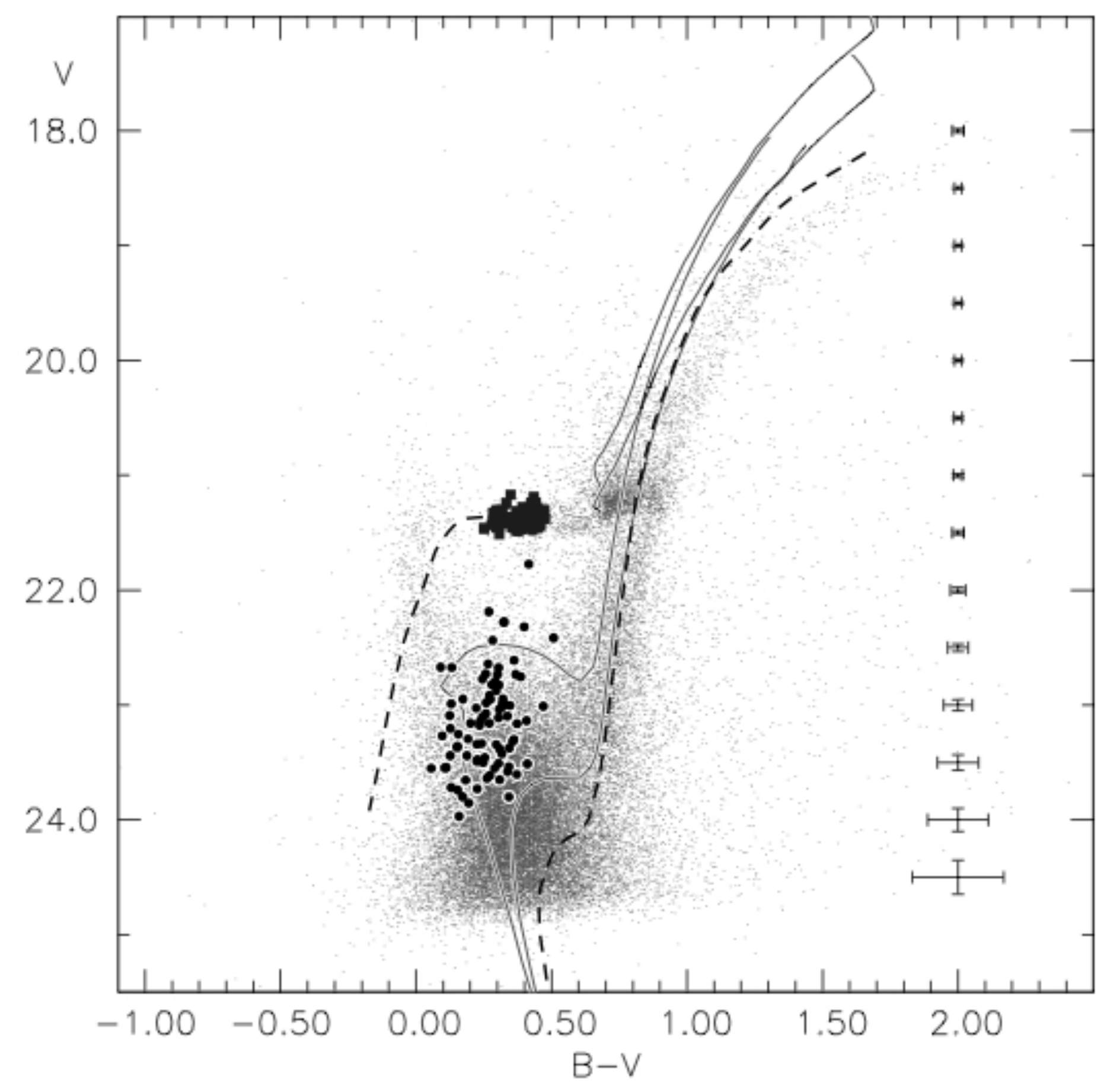}
\par\end{centering}

\caption{The colour-magnitude diagram of the Fornax dwarf galaxy\index{Fornax dwarf galaxy}. Highlighted
with heavy squares are the RR Lyrae\index{RR Lyrae star}, while the sample of 85 SX Phe
variables are highlighted with heavy filled circles. This figure is
reproduced from \cite{Poretti2008} with permission by the AAS.
\label{momfig:Poretti_Sx_phoe}}
\end{figure}

Poretti et al. \cite{Poretti2008} presented an intensive survey of the Fornax dwarf\index{Fornax dwarf galaxy}
(a galaxy known to harbour recent star formation activity and a predominately
intermediate age population). Figure$\,$\ref{momfig:Poretti_Sx_phoe}
shows how the group of 85 SX Phe variables is coincident with the
expected BSS location and, at the same time, immersed within the blue
plume of young stars. Examining the period--luminosity relation for
the SX Phe sample, Poretti et al. \cite{Poretti2008} conclude that the observed
scatter exceeds their observational errors and propose a physical
rather than an instrumental origin for this scatter. Using the period-luminosity
plane, they reported the first identification of a peculiar group
of \emph{sub-luminous} extra-Galactic SX Phe variables. The presence
of the sub-luminous SX Phe was confirmed by {Cohen} \& {Sarajedini} \cite{Cohen2012b} in
the Carina dwarf\index{Carina dwarf galaxy}, NGC 2419\index{NGC 2419} and $\omega$ Cen\index{$\omega$ Centauri}. Poretti et al. \cite{Poretti2008} cautiously
speculate that the \emph{sub-luminous} variables could be the results
of merging of a close binary system, and present arguments in favour
for this scenario. Clearly, more observations are needed to confirm
this scenario. However, this may open a new field where BSS can be
disentangled (via their variability signature) in star forming dwarf
galaxies.

\section{The Progeny of BSS\label{momsec:The-Progeny-of-BSS}}

\begin{figure}
\begin{centering}
\includegraphics[width=59mm]{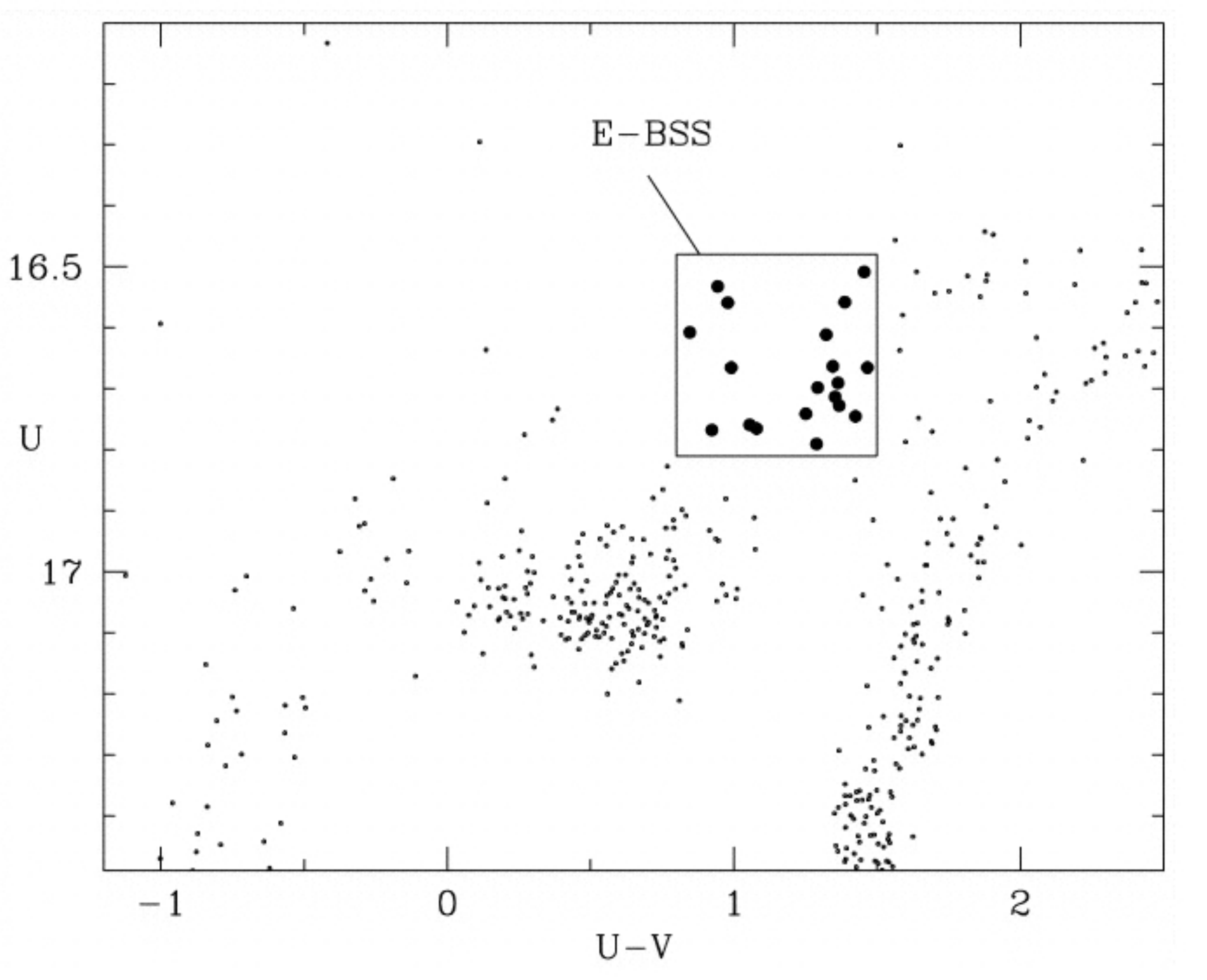}\includegraphics[width=59mm]{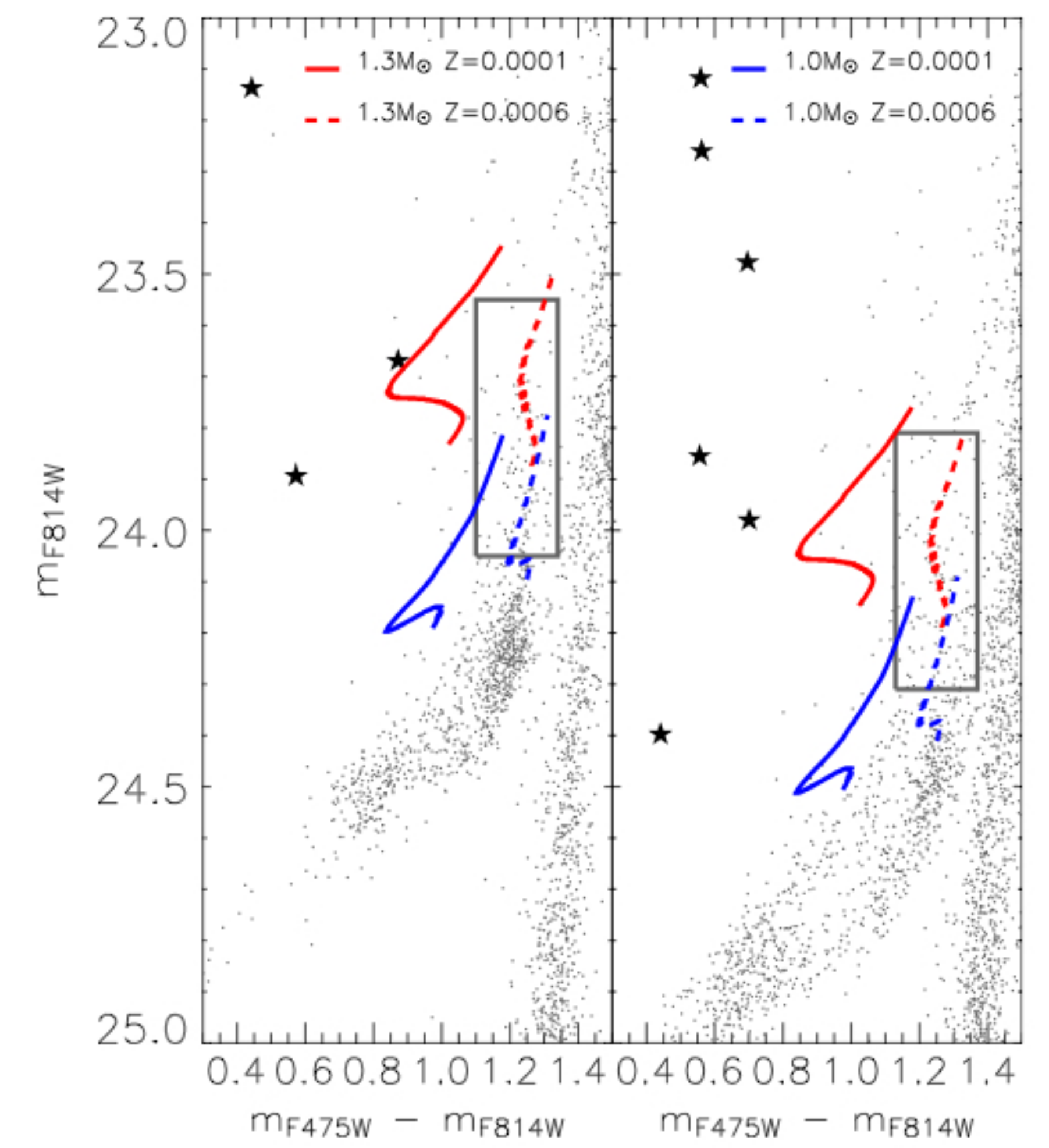}
\par\end{centering}

\caption{The left panel displays the \emph{HST} ultraviolet $U,$ ($U-V$) diagram of
M80\index{M80} where the progeny of BSS are identified within the open box. The right
panel displays the optical diagram of the Cetus\index{Cetus dwarf galaxy} and Tucana dwarf galaxies\index{Tucana dwarf galaxy},
where the boxes also highlight the E-BSS population. The filled stars
are the identified Anomalous Cepheid variables\index{Anomalous Cepheid variable}, detected by \cite{Bernard2009c}.
The left and right figures are reproduced from \cite{Monelli2012}, and \cite{Ferraro1999a}, respectively, with permission by the
AAS. \label{momfig:Ferraro-M80_Evolved-BSS}}
\end{figure}

During their core-helium burning phase, the progeny of the BSS population
(i.e. evolved-BSS, hereafter E-BSS) are expected to pile up in a particular
location in the colour-magnitude diagram: bluer than the red giant
branch and brighter than normal horizontal branch stars. {Renzini} \& {Fusi Pecci} \cite{Renzini1988}
were the first to suggest the presence of E-BSS. Interestingly, the
occurrence of E-BSS is independent of the formation channel of the
BSS \cite{Sills2009a}. Ferraro et al. \cite{Ferraro1999a} performed a systematic
search for E-BSS, and Fig.\ref{momfig:Ferraro-M80_Evolved-BSS} shows
their ultraviolet\index{ultraviolet} diagram of M80\index{M80} where the E-BSS population is clearly
present and highlighted by the open box. They derive $N_{BSS}/N_{E-BSS}=16.$ 

The right panel of Fig.\ref{momfig:Ferraro-M80_Evolved-BSS} shows the
detection of a vertical extension in correspondence of the red HB
region in the Cetus and Tucana dwarf galaxies. In the context of dwarf
galaxies, this sequence is usually referred to as \emph{vertical-clump}
\cite{Gallart1999c}: helium-burning stars of few hundred Myr
to $1-2$ Gyr old population. Monelli et al. \cite{Monelli2010a} use their best
reconstruction of the Cetus star formation history, and derive a $N_{BSS}/N_{E-BSS}=12$
and $10$ upon simulated and observed diagrams, respectively. The
agreement between the empirical and synthetic values for the Cetus
dwarf, and the value derived by \cite{Ferraro1999a} all point to
a correct identification of the\emph{ same} evolved BSS population
in the 2 different systems. We repeated the exercise for the LeoII
dwarf galaxy and derive $N_{BSS}/N_{E-BSS}=9$, again indicating a
correct identification of E-BSS in dwarf galaxies.

\textbf{\emph{The case of Anomalous Cepheids}}\index{Anomalous Cepheid variable}: dwarf galaxies (whether
\emph{purely ancient} systems like Ursa Minor or intermediate-age
ones like LeoI) are known to host a peculiar class of Cepheid variables
known as Anomalous Cepheids (ACs). RR Lyrae and ACs are core helium-burning
pulsational variables passing through the instability strip and co-exist
in low-metallicity systems. However, ACs are $2-3$ times more massive
than RR Lyrae \cite{Renzini1977b,Fiorentino2006a}.
The origin of ACs is sought as either due to: (i) relatively young
($\sim1-6$ Gyr) single stars; or (ii) the progeny of BSS, formed
through mass transfer in primordial ($10$ Gyr) binary systems.
The early study of {Renzini}, {Mengel} \& {Sweigart} \cite{Renzini1977b} shows how the chances of
survival of such primordial binaries are strictly correlated with
the stellar density of the parent system: ACs are easily destroyed
via dynamical encounters in high density environments like globular
clusters. On the other hand, the low stellar densities of dwarf galaxies
offer a peaceful environment and allow the system to evolve. This
1977 prediction is proven correct by the continuous and successful
surveys of ACs in dwarf galaxies --- e.g. Bernard et al. \cite{Bernard2009c} for
the detection of $8$ and $6$ ACs in Cetus and Tucana, and Fiorentino et al. \cite{Fiorentino2012b}
for the detection of $51$ ACs in Leo$\,$I. Most interestingly, the
prediction of the almost \emph{lack} of ACs in globular clusters still
holds as well. The exception being the detection of a \emph{single}
AC detected in NGC 5466\index{NGC 5466} \cite{Zinn1976}, which \emph{is} a
low-density cluster.

\begin{figure}
\sidecaption
\includegraphics[width=75mm]{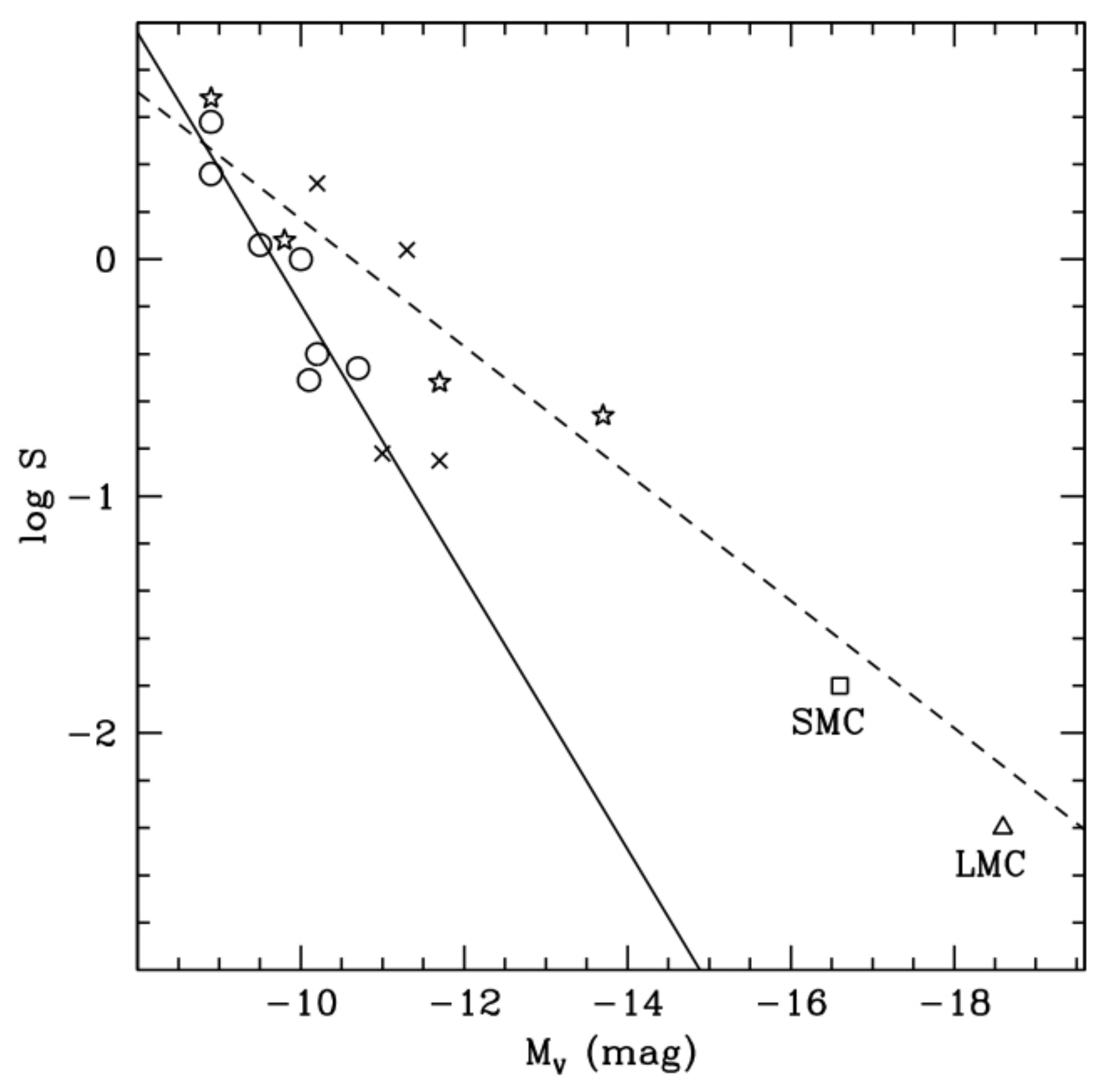}
\caption{The Anomalous Cepheid frequency ($log\, S$, per $10^{5}L_{\odot}$)
as a function of the M$_{V}$ of the parent dwarf galaxy (i.e. the
\emph{ACs frequency--M$_{V}$ anti-correlation}). Open circles highlight
purely ancient galaxies while starred symbols indicate galaxies with
large intermediate-age populations. The solid line shows an excellent
fit to only the ancient systems (included in this review). This figure
is reproduced from \cite{Fiorentino2012f} with permission by
Astronomy \& Astrophysics, \tiny{\copyright} ESO. \label{momfig:Fiorentino-AC-Mv}}
\end{figure}

Fiorentino \& {Monelli} \cite{Fiorentino2012f} complemented the earlier studies by  Mateo et al. \cite{Mateo1995c}
and Pritzl et al. \cite{Pritzl2002c} and reinvestigated the correlation between
the ACs frequency and the absolute visual luminosity of the parent
galaxy. Figure$\,$\ref{momfig:Fiorentino-AC-Mv} displays the Anomalous
Cepheid frequency ($\log\, S$, per $10^{5}L_{\odot}$) as a function
of the M$_{V}$ of the parent dwarf galaxy. Open circles show purely
ancient galaxies while starred symbols indicate galaxies with large
intermediate-age populations. The sample of \cite{Fiorentino2012b} 
of ancient galaxies includes: Ursa Minor\index{Ursa Minor dwarf galaxy}, Draco\index{Draco dwarf galaxy}, Sculptor\index{Sculptor dwarf galaxy}, Leo$\,$II\index{Leo$\,$II dwarf galaxy},
Sextans\index{Sextans dwarf galaxy}, Tucana\index{Tucana dwarf galaxy} and Cetus\index{Cetus dwarf galaxy} (all of which are included in the present
BSS study) confirm that these possess a genuine BSS population. Figure$\,$\ref{momfig:Fiorentino-AC-Mv}
shows that for these galaxies there exists a clear anti-correlation
between the ACs frequency and the luminosity of the parent galaxy.
The solid line shows the fit to \emph{only} the ancient systems while
the dashed line is the fit for the intermediate-age galaxies. The
interpretation of the apparent ACs frequency--M$_{V}$ anti-correlations
for the two groups is not straightforward. However, intermediate-age
galaxies systematically show higher ACs frequencies, and this goes
along with the fact that these systems can host ACs originating from
2 different channels (i.e. recently formed young stars as well as
evolving primordial binaries). {Fiorentino} \& {Monelli} \cite{Fiorentino2012f} note, however,
that for their sample of ancient systems one expects to find ACs originating
\emph{only} through the binary channel, and therefore imply that binary
systems have a higher chance of survival in low-mass galaxies.

\section{Conclusion}

During the workshop, George Preston proposed us a nice definition of blue straggler stars:
\begin{quotation}
BSS are a subset
of the interacting-binary field stars whose end products
are hotter than the main sequence turn-off of the parent stellar population.
\end{quotation}

In his 1993 review L. Stryker \cite{Stryker1993} concludes 
\begin{quotation}
This
makes forming conclusion about the origin of BS in dwarf galaxies
difficult, if not impossible, at the present time. 
\end{quotation}

Twenty years
after, I believe that the recent imaging surveys of these distant
systems have allowed us to get a deeper insight on the general properties
and origin of their BSS population. Indeed, and thanks to the new
generations of ground and space-based telescopes\index{telescope}, the star formation
and chemical enrichment history for the majority of Local Group galaxies\index{Local Group galaxy}
have been derived. This has allowed us to scrutinise and filter-out
those galaxies with hints of recent star formation activity. Consequently,
clean samples of ``ancient'' galaxies are available to address the
\emph{BSS-young stars} ambiguity in a reasonable manner. 

Our current understanding of the possible BSS formation channels are
mass-transfer binaries, collisions and mergers, and all three channels
must manifest themselves in globular clusters\index{globular cluster}. 
The central peak of
the BSS radial distribution is \emph{still} best explained as a signature
of collisional BSS. Similarly, the segregation of the brighter BSS
in the cluster centres is also attributed to collisional BSS. Assuming that the previous two interpretations are true, one can confidently
affirm that both these collisional BSS \emph{``signatures''} are
lacking in dwarf galaxies. This proves the long-suspected prediction
that BSS in dwarf galaxies form preferentially through mass-transfer
binaries, as is the case for the Galactic Halo\index{Halo}.

Figure$\,$\ref{momfig:Freq-vs-Mv} shows that the lowest luminosity
dwarf galaxies with $-8.0\leq M_{V}\leq-5.0$ have BSS frequencies
that are in perfect agreement with that observed in Galactic open
clusters\index{open
cluster}. Evidence that the lowest-luminosity globular clusters share
the same BSS frequency as in open clusters \emph{and} dwarf galaxies
allow us to suggest that there exists an \emph{empirical upper limit}
to the BSS frequency in \emph{any} given stellar system. Figure$\,$\ref{momfig:Freq-vs-Mv}
also shows that the BSS frequency in dwarf galaxies with M$_{V}\approx-9.0$
is \emph{higher} than that of globular clusters at the same luminosity.
This is in perfect agreement with the notion that higher density environments,
with their higher collisional rates, perturb the evolution of mass transfer
in primordial binaries and end up being BSS deficient. Dwarf galaxies
on the other hand offer a more friendly environment for mass-transfer
binaries and tend to preserve their initial binary population. The
recent derivation of the BSS frequency in the Galactic Bulge\index{Bulge}, and
arguments presented for the BSS frequency in the Galactic Halo, seem
to provide more evidence for the dwarf galaxies \emph{BSS frequency--M$_{V}$}
\emph{anti-correlation}. The mechanism responsible for the decreasing
BSS frequency with increasing mass remains unclear, and awaits further
confirmation of the \emph{anti-correlation. } Interestingly, a \emph{tight
} and \emph{ strong} anti-correlation (Fig.\ref{momfig:Fiorentino-AC-Mv})
has been observed between the frequency of Anomalous Cepheids (a progeny
of BSS) and M$_{V}$ for ancient dwarf galaxies with $-11.0\leq M_{V}\leq-9.0$. 

All together, I believe that the past few years have been very fruitful
for the study of BSS in dwarf galaxies, and the best (deeper imaging
and kinematic and chemical surveys) are yet to come. 

\begin{acknowledgement}
I would like to thank S. Zaggia for valuable comments. H. Boffin and
G. Carraro are warmly thanked for organising a wonderful workshop. 
\end{acknowledgement}

\backmatter
\printindex



\begin{thebibliography}{79}
\expandafter\ifx\csname natexlab\endcsname\relax\def\natexlab#1{#1}\fi

\bibitem{Aparicio2001c}
{Aparicio}, A., {Carrera}, R.,  {Mart{\'{\i}}nez-Delgado}, D.:  AJ {\bf 122},
  2524 (2001)

\bibitem{Beccari2008}
{Beccari}, G., {Pulone}, L., {Ferraro}, F.~R., {et~al.}: MemSAI {\bf 79}, 360 (2008)

\bibitem{Beccari2011}
{Beccari}, G., {Sollima}, A., {Ferraro}, F.~R., {et~al.}: ApJL {\bf 737}, 3 (2011)

\bibitem{Belokurov2006c}
{Belokurov}, V., {Zucker}, D.~B., {Evans}, N.~W., {et~al.}: ApJL {\bf 647},
  111 (2006)

\bibitem{Bernard2009c}
{Bernard}, E.~J., {Monelli}, M., {Gallart}, C., {et~al.}: ApJ {\bf 699}, 1742 (2009)

\bibitem{Blumenthal1984d}
{Blumenthal}, G.~R., {Faber}, S.~M., {Primack}, J.~R.,  {Rees}, M.~J.:
 Nature {\bf 311}, 517 (1984)

\bibitem{Bono2010e}
{Bono}, G., {Stetson}, P.~B., {Walker}, A.~R., {et~al.}: PASP {\bf 122}, 651 (2010)

\bibitem{Brown2008}
{Brown}, T.~M., {Beaton}, R., {Chiba}, M., {et~al.}: ApJL {\bf 685}, 121 (2008)

\bibitem{Carrera2002e}
{Carrera}, R., {Aparicio}, A., {Mart{\'{\i}}nez-Delgado}, D., \&
  {Alonso-Garc{\'{\i}}a}, J.: AJ {\bf 123}, 3199 (2002)

\bibitem{Carretta2009f}
{Carretta}, E., {Bragaglia}, A., {Gratton}, R.~G., {et~al.}: A\&A {\bf 505},
  117 (2009)

\bibitem{Clarkson2011c}
{Clarkson}, W.~I., {Sahu}, K.~C., {Anderson}, J., {et~al.}: ApJ {\bf 735}, 37 (2011)

\bibitem{Cohen2012b}
{Cohen}, R.~E.  {Sarajedini}, A.: {\bf 419}, 342 (2012)

\bibitem{Davies2004a}
{Davies}, M.~B., {Piotto}, G.,  {de Angeli}, F.: {\bf 349}, 129 (2004)

\bibitem{2006A&A...459..489D}
{de Marchi}, F., {de Angeli}, F., {Piotto}, G., {Carraro}, G.,  {Davies},
  M.~B.: A\&A {\bf 459}, 489 (2006)

\bibitem{Ferraro1997e}
{Ferraro}, F.~R., {Paltrinieri}, B., {Fusi Pecci}, F., {et~al.}: A\&A {\bf 324}, 915 (1997)

\bibitem{Ferraro1999a}
{Ferraro}, F.~R., {Paltrinieri}, B., {Rood}, R.~T.,  {Dorman}, B.: ApJ {\bf 
  522}, 983 (1999)

\bibitem{Ferraro2006a}
{Ferraro}, F.~R., {Sollima}, A., {Rood}, R.~T., {et~al.}: ApJ {\bf 638}, 433 (2006)

\bibitem{Fiorentino2006a}
{Fiorentino}, G., {Limongi}, M., {Caputo}, F.,  {Marconi}, M.: A\&A {\bf 460}, 155 (2006)

\bibitem{Fiorentino2012f}
{Fiorentino}, G.  {Monelli}, M.: A\&A {\bf 540}, A102 (2012)

\bibitem{Fiorentino2012b}
{Fiorentino}, G., {Stetson}, P.~B., {Monelli}, M., {et~al.}: ApJL {\bf 759},
  12 (2012)

\bibitem{FusiPecci1992}
{Fusi Pecci}, F., {Ferraro}, F.~R., {Corsi}, C.~E., {Cacciari}, C., \&
  {Buonanno}, R.: AJ {\bf 104}, 1831 (1992)

\bibitem{Gallart1999c}
{Gallart}, C., {Freedman}, W.~L., {Mateo}, M., {et~al.}: ApJ {\bf 514}, 665 (1999)

\bibitem{Gilmore2007cm}
{Gilmore}, G., {Wilkinson}, M.~I., {Wyse}, R.~F.~G., {et~al.}: ApJ {\bf 663},
  948 (2007)

\bibitem{Girardi2005}
{Girardi}, L., {Groenewegen}, M.~A.~T., {Hatziminaoglou}, E.,  {da Costa}, L.: A\&A {\bf 436}, 895 (2005)

\bibitem{Held2005}
{Held}, E.~V.: , in IAU Colloq. 198: Near-fields cosmology with dwarf
  elliptical galaxies, p. 11 (2005)

\bibitem{Held2000}
{Held}, E.~V., {Saviane}, I., {Momany}, Y.,  {Carraro}, G.: ApJL {\bf 530},
  85 (2000)

\bibitem{Hills1976d}
{Hills}, J.~G.  {Day}, C.~A.:  ApJL {\bf 17}, 87 (1976)

\bibitem{Holtzman2006d}
{Holtzman}, J.~A., {Afonso}, C.,  {Dolphin}, A.: ApJS {\bf 166}, 534 (2006)

\bibitem{Hurley-Keller1998}
{Hurley-Keller}, D., {Mateo}, M.,  {Nemec}, J.: AJ {\bf 115}, 1840 (1998)

\bibitem{Keller2012am}
{Keller}, S.~C., {Skymapper Team},  {Aegis Team}: in Galactic Archaeology:
  Near-Field Cosmology and the Formation of the Milky Way, ASPC 458, p. 409 (2012)

\bibitem{Lanzoni2007}
{Lanzoni}, B., {Dalessandro}, E., {Ferraro}, F.~R., {et~al.}:
 ApJ {\bf 663}, 267 (2007a)

\bibitem{Lanzoni2007a}
{Lanzoni}, B., {Sanna}, N., {Ferraro}, F.~R., {et~al.}:
  ApJ {\bf 663}, 1040 (2007b)

\bibitem{Lee2003u}
{Lee}, M.~G., {Park}, H.~S., {Park}, J.-H., {et~al.}: AJ {\bf 126}, 2840 (2003)

\bibitem{Mandushev1997}
{Mandushev}, G.~I., {Fahlman}, G.~G., {Richer}, H.~B.,  {Thompson}, I.~B.: AJ {\bf 114}, 1060 (1997)

\bibitem{Mapelli2009d}
{Mapelli}, M., {Ripamonti}, E., {Battaglia}, G., {et~al.}: MNRAS {\bf 396},
  1771 (2009)

\bibitem{Mapelli2007b}
{Mapelli}, M., {Ripamonti}, E., {Tolstoy}, E., {et~al.}: MNRAS {\bf 380}, 1127 (2007)

\bibitem{Mapelli2004}
{Mapelli}, M., {Sigurdsson}, S., {Colpi}, M., {et~al.}: ApJL {\bf 605}, 29 (2004)

\bibitem{Mapelli2006c}
{Mapelli}, M., {Sigurdsson}, S., {Ferraro}, F.~R., {et~al.}: MNRAS {\bf  373},
  361 (2006)

\bibitem{Marin-Franch2009}
{Mar{\'{\i}}n-Franch}, A., {Aparicio}, A., {Piotto}, G., {et~al.}: ApJ {\bf 694}, 1498 (2009)

\bibitem{Martin2008}
{Martin}, N.~F., {Coleman}, M.~G., {De Jong}, J.~T.~A., {et~al.}: ApJL {\bf 672}, 13 (2008)

\bibitem{Mateo1995c}
{Mateo}, M., {Fischer}, P.,  {Krzeminski}, W.:  AJ {\bf 110}, 2166 (1995)

\bibitem{Mateo1998d}
{Mateo}, M., {Hurley-Keller}, D.,  {Nemec}, J.:  AJ {\bf 115}, 1856 (1998)

\bibitem{Mateo1991a}
{Mateo}, M., {Nemec}, J., {Irwin}, M.,  {McMahon}, R.: AJ {\bf 101}, 892 (1991)

\bibitem{Mateo1998}
{Mateo}, M.~L.:  ARA\&A {\bf 36}, 435 (1998)

\bibitem{McConnachie2012a}
{McConnachie}, A.~W.: AJ {\bf 144}, 4 (2012)

\bibitem{McCrea1964a}
{McCrea}, W.~H.: MNRAS {\bf 128}, 147 (1964)

\bibitem{Milone2013d}
{Milone}, A.~P., {Marino}, A.~F., {Piotto}, G., {et~al.}: ApJ {\bf 767}, 120 (2013)

\bibitem{Misgeld2011}
{Misgeld}, I.  {Hilker}, M.: MNRAS {\bf 414}, 3699 (2011)

\bibitem{Momany2003b}
{Momany}, Y., {Cassisi}, S., {Piotto}, G., {et~al.}: A\&A {\bf 407}, 303 (2003)

\bibitem{Momany2007}
{Momany}, Y., {Held}, E.~V., {Saviane}, I., {et~al.}: A\&A {\bf 468}, 973 (2007)

\bibitem{Monaco2003a}
{Monaco}, L., {Bellazzini}, M., {Ferraro}, F.~R.,  {Pancino}, E.:  ApJL {\bf 597}, 25 (2003)

\bibitem{Monelli2012}
{Monelli}, M., {Cassisi}, S., {Mapelli}, M., {et~al.}: ApJ {\bf 744}, 157 (2012)

\bibitem{Monelli2010a}
{Monelli}, M., {Hidalgo}, S.~L., {Stetson}, P.~B., {et~al.}: ApJ {\bf 720},
  1225 (2010)

\bibitem{Monkman2006}
{Monkman}, E., {Sills}, A., {Howell}, J., {et~al.}: ApJ {\bf 650}, 195 (2006)

\bibitem{Niederste-Ostholt2009}
{Niederste-Ostholt}, M., {Belokurov}, V., {Evans}, N.~W., {et~al.}:
MNRAS  {\bf 398}, 1771 (2009)

\bibitem{Piotto2004c}
{Piotto}, G., {De Angeli}, F., {King}, I.~R., {et~al.}: ApJL {\bf 604}, 109 (2004)

\bibitem{Piotto2002}
{Piotto}, G., {de Angeli}, F., {Recio Blanco}, A., {et~al.}: in
  Observed HR
  Diagrams and Stellar Evolution, ASPC 274, p. 282 (2002)

\bibitem{Poretti1999}
{Poretti}, E.:  A\&A {\bf 343}, 385 (1999)

\bibitem{Poretti2008}
{Poretti}, E., {Clementini}, G., {Held}, E.~V., {et~al.}: ApJ {\bf 685}, 947 (2008)

\bibitem{Preston2000e}
{Preston}, G.~W.  {Sneden}, C.: AJ {\bf 120}, 1014 (2000)

\bibitem{Pritzl2002c}
{Pritzl}, B.~J., {Armandroff}, T.~E., {Jacoby}, G.~H.,  {Da Costa}, G.~S.: AJ {\bf 124}, 1464 (2002)

\bibitem{Renzini2013}
{Renzini}, A.: MmSAI {\bf
84}, 162 
(2013)

\bibitem{Renzini1988}
{Renzini}, A.  {Fusi Pecci}, F.: ARA\&A {\bf 26}, 199 (1988)

\bibitem{Renzini1977b}
{Renzini}, A., {Mengel}, J.~G.,  {Sweigart}, A.~V.:  A\&A {\bf 56}, 369 (1977)

\bibitem{Rizzi2003e}
{Rizzi}, L., {Held}, E.~V., {Bertelli}, G.,  {Saviane}, I.:
  ApJL {\bf 589}, 85 (2003a)

\bibitem{Rizzi2003}
{Rizzi}, L., {Held}, E.~V., {Momany}, Y., {et~al.}: MmSAI {\bf 74}, 510 (2003b)

\bibitem{YoSandage1953}
{Sandage}, A.~R.: AJ {\bf 58}, 61 (1953)

\bibitem{Santana2012}
{Santana}, F.~A., {Mu{\~n}oz}, R.~R., {Geha}, M., {et~al.}: in
  Galactic
  Archaeology: Near-Field Cosmology and the Formation of the Milky Way, ASPC 458, p. 339 (2012)

\bibitem{Saviane2000d}
{Saviane}, I., {Held}, E.~V.,  {Bertelli}, G.: A\&A {\bf 355}, 56 (2000)

\bibitem{Saviane2003}
{Saviane}, I., {Rosenberg}, A., {Aparicio}, A.,  {Piotto}, G.: in
  New Horizons
  in Globular Cluster, ASPC 296, p. 402 (2003)

\bibitem{Sills2009a}
{Sills}, A., {Karakas}, A.,  {Lattanzio}, J.: ApJ {\bf 692}, 1411 (2009)

\bibitem{Sollima2008c}
{Sollima}, A., {Lanzoni}, B., {Beccari}, G., {Ferraro}, F.~R.,  {Fusi Pecci},
  F.: A\&A {\bf 481}, 701 (2008)

\bibitem{Stryker1993}
{Stryker}, L.~L.: PASP {\bf 105}, 1081 (1993)

\bibitem{White1978e}
{White}, S.~D.~M.  {Rees}, M.~J.: MNRAS {\bf 183}, 341 (1978)

\bibitem{Willman2005a}
{Willman}, B., {Dalcanton}, J.~J., {Martinez-Delgado}, D., {et~al.}:
  ApJL {\bf 626}, 85 (2005)

\bibitem{Willman2012}
{Willman}, B.  {Strader}, J.: AJ {\bf 144}, 76 (2012)

\bibitem{Zaggia1997}
{Zaggia}, S.~R., {Piotto}, G.,  {Capaccioli}, M.: A\&A {\bf 327}, 1004 (1997)

\bibitem{Zhao2012bl}
{Zhao}, Z., {Okamoto}, S., {Arimoto}, N., {Aoki}, W.,  {Kodama}, T.: in
  Galactic
  Archaeology: Near-Field Cosmology and the Formation of the Milky Way, ASPC 458,  p. 349 (2012)

\bibitem{Zinn1976}
{Zinn}, R.  {Dahn}, C.~C.: AJ {\bf 81}, 527 (1976)

\end{thebibliography}
\end{document}